\documentclass[]{aastex631}

\usepackage{amsmath}
\usepackage{ulem}
\usepackage{appendix}

\begin{document}

\title{Evolution of the Shock Properties of the 2023 March 13 Event from In-Situ and Remote-Sensing Data}

\correspondingauthor{Federica Chiappetta}
\email{federica.chiappetta@unical.it}

\author[0000-0001-7221-1382]{Federica Chiappetta}
\affiliation{Dipartimento di Fisica, Universit\`a della Calabria, Rende, Italy \\}

\author[0000-0003-2566-2820]{Giuseppe Nisticò}
\affiliation{Dipartimento di Fisica, Universit\`a della Calabria, Rende, Italy \\}

\author[0009-0004-4989-3484]{Massimo Chimenti}
\affiliation{Dipartimento di Fisica, Universit\`a della Calabria, Rende, Italy \\}

\author[0000-0002-7653-9147]{Andrea Larosa}
\affiliation{CNR, Institute for Plasma Science and Technology (ISTP), Bari, Italy \\}

\author[0000-0002-5554-8765]{Francesco Malara}
\affiliation{Dipartimento di Fisica, Universit\`a della Calabria, Rende, Italy \\}

\author[0000-0002-5272-5404]{Francesco Pucci}
\affiliation{CNR, Institute for Plasma Science and Technology (ISTP), Bari, Italy \\}

\author[0000-0002-5981-7758]{Luca Sorriso-Valvo}
\affiliation{CNR, Institute for Plasma Science and Technology (ISTP), Bari, Italy \\}
\affiliation{KTH, Division of Space and Plasma Physics, Stockholm, Sweden \\}

\author[0000-0002-9207-2647]{Gaetano Zimbardo}
\affiliation{Dipartimento di Fisica, Universit\`a della Calabria, Rende, Italy \\}

\author[0000-0002-8399-3268]{Silvia Perri}
\affiliation{Dipartimento di Fisica, Universit\`a della Calabria, Rende, Italy \\}

\begin{abstract}
Shocks driven by coronal mass ejections (CMEs) are the most powerful accelerators of gradual solar energetic particles (SEPs) in the inner heliosphere. On 2023 March 13, a halo CME, as seen from the Solar Heliospheric Observatory (SoHO) and the Sun TErrestrial Relations Observatory (STEREO), gave rise to a strong SEP event.
In this work, we aim to analyze this CME-driven shock from multiple spacecraft, using both remote sensing observations from STEREO-A/COR2 and in-situ data from Parker Solar Probe (PSP), Solar Orbiter (SolO), and Wind.
In order to determine its direction of propagation and kinematic properties, we model the shock geometry using STEREO-A/COR2 and SoHO/LASCO/C3 observations as an expanding ellipsoid. The density compression ratio of the shock is determined by fitting the brightness profile from the coronagraphic images with that obtained from raytracing simulations of a double-Gaussian shock density profile. We compare physical quantities such as compression ratio and Alfv\'enic Mach number derived from remote sensing observations with in-situ measurements by PSP, SolO, STEREO-A, and Wind. From STEREO-A/COR2, we determine the compression ratio around the entire shock front in the corona, finding significant non-homogeneities that can impact the values found during in-situ crossings.
Following the evolution of the parameters characterizing the CME from the source to space, we find that closer to the Sun, both the gas compression ratio and the Alfv\'enic Mach number remain almost constant, while they increase at larger radial distances. This indicates a non-trivial evolution of the shock parameters during its journey through the interplanetary space.
\end{abstract}

\keywords{Shocks --- Solar corona --- Solar coronal mass ejection --- Solar energetic particles}

\section{Introduction} \label{sec:intro}
Solar energetic particles (SEPs) represent a natural hazard in the near-Earth environment for spacecraft components and payload, for communication networks, and for astronauts during space missions. Thus, predicting the arrival time of such a flux of energetic particles towards the Earth has become a science top priority in the last years, also in the context of future space missions to the Moon and Mars. 
SEPs are mainly accelerated at shocks formed in front of expanding fast coronal mass ejections \citep[CMEs; ][]{reames_1999,desai-giacalone_2016,kilpua_2017}, the so-called gradual event, characterized by a ``sharp" onset of the time profile of SEP fluxes and a slow decaying phase. As detected in-situ, SEPs events can also be characterized by an enhancement of their flux at the shock during the satellite crossing \citep{chiappetta_2021,reames_2021}. To achieve a significant leap forward in the predictive capabilities of SEP events, it is important to investigate the properties of coronal shocks associated with CMEs, as observed both at the solar source and in interplanetary space. First observations of coronal shocks with remote sensing instruments were provided by the Large Angle and Spectrometer Coronagraph (LASCO) of the Solar and Heliospheric Observatory (SoHO). A shock appears as a very faint emission in coronagraphs in front of the leading edge of expanding CMEs \citep{ontiveros-vourlidas_2009,kwon-vourlidas_2017}. Shocks have also been observed in EUV in the low solar corona, in association with expanding global coronal waves \citep{ma_2011, temmer_2013}, manifesting also a dome-like shape \citep{mann2023}. 

A key parameter to physically characterize a CME-driven shock is the density compression ratio. Estimation of this quantity, defined as the ratio between the downstream and upstream electron densities, is important to describe shocks during the initial phase of their expansion in the high corona from 4 to 15 solar radii and to estimate the energetic particle spectral index. Studying the evolution of the density compression ratio is essential since the association between the CME-driven shocks and the SEP population in the corona is increasingly evident \citep{rouillard_2014,rouillard_2016,lario_2016}. 
Inferring the compression ratio from remote-sensing data is not a trivial procedure and requires proper modeling of the shock.
Coronagraphic images return the Thomson-scattered emission of free electrons in coronal structures (e.g., streamers, CMEs, etc.) integrated along the line-of-sight (LoS). Multiple observations of the same event from different points of view may offer a better estimation of the electron density distribution in coronal structures. Thanks to different spacecraft with multi-viewpoints, it is possible to determine the 3D geometric structure of the shock front using an ellipsoid model \citep{kwon_2014,kwon_2015} and to represent the density profile with a double Gaussian-like shape \citep{kwon-vourlidas_2018}. In this scenario, the radial profile of the electron density for a quiet corona is estimated by inverting the polarized brightness ($pB$) measurements \citep{vandeHulst_1950} or the total brightness images \citep{hayes_2001}. Thomson scattering of photospheric light off free electrons produces the polarized component of the white-light coronal emission, forming the so-called K-corona. Then, the $pB$ signal carries information about the electron density distribution in the corona, since the brightness contributions from the F corona, arising from interplanetary dust, are automatically excluded. The radial evolution of the electron density in the quiet corona can be estimated with different models using a polynomial function \citep{saito_1977,leblanc_1998,hayes_2001}. 

In this work, we provide a thorough analysis of the CME-driven shock that occurred on 2023 March 13, classified as a halo CME in the LASCO catalog\footnote{\url{https://cdaw.gsfc.nasa.gov/CME_list/UNIVERSAL_ver2/2023_03/univ2023_03.html}}. The CME expansion was observed by multiple spacecraft, providing both remote sensing observations and direct in-situ measurements. From the coronagraphic images, it is evident that the shock front is visible around the entire solar disk. In fact, \citet{dresing_2025}, thanks to the opportunity of multi-point observations of such an event, were able to identify it as a widespread SEP event, since almost all the satellites detecting the increase in the energetic particle fluxes are longitudinally far separated from the probable active region and provided an explanation of the circumsolar interplanetary shock. The paper is organized as follows. Section \ref{sec:data} describes the analyzed event, from remote sensing observations to in-situ data. In Section \ref{sec:model} we present the shock model and the raytracing simulation of the Thomson scattered emission. In particular, we use an ellipsoid model to represent the propagation of the shock front as observed by STEREO-A/COR2 and SoHO/LASCO/C3 and we produce synthetic images using the raytracing simulations of the Thomson scattered emission applied to a datacube of density \citep{thernisien_2006,thernisien_2011}. Section \ref{sec:results} discusses data analysis and the results obtained from the source to space, comparing the theoretical model with the observations and deriving an estimation of the physical quantities associated with the shock, e.g. the compression ratio and the Alfv\'enic Mach number. Finally, discussion and conclusions are given in Section \ref{sec:conclusions}.

\section{Observations and data}
\label{sec:data}
\begin{figure}
    \centering
    \includegraphics[width=0.45\linewidth]{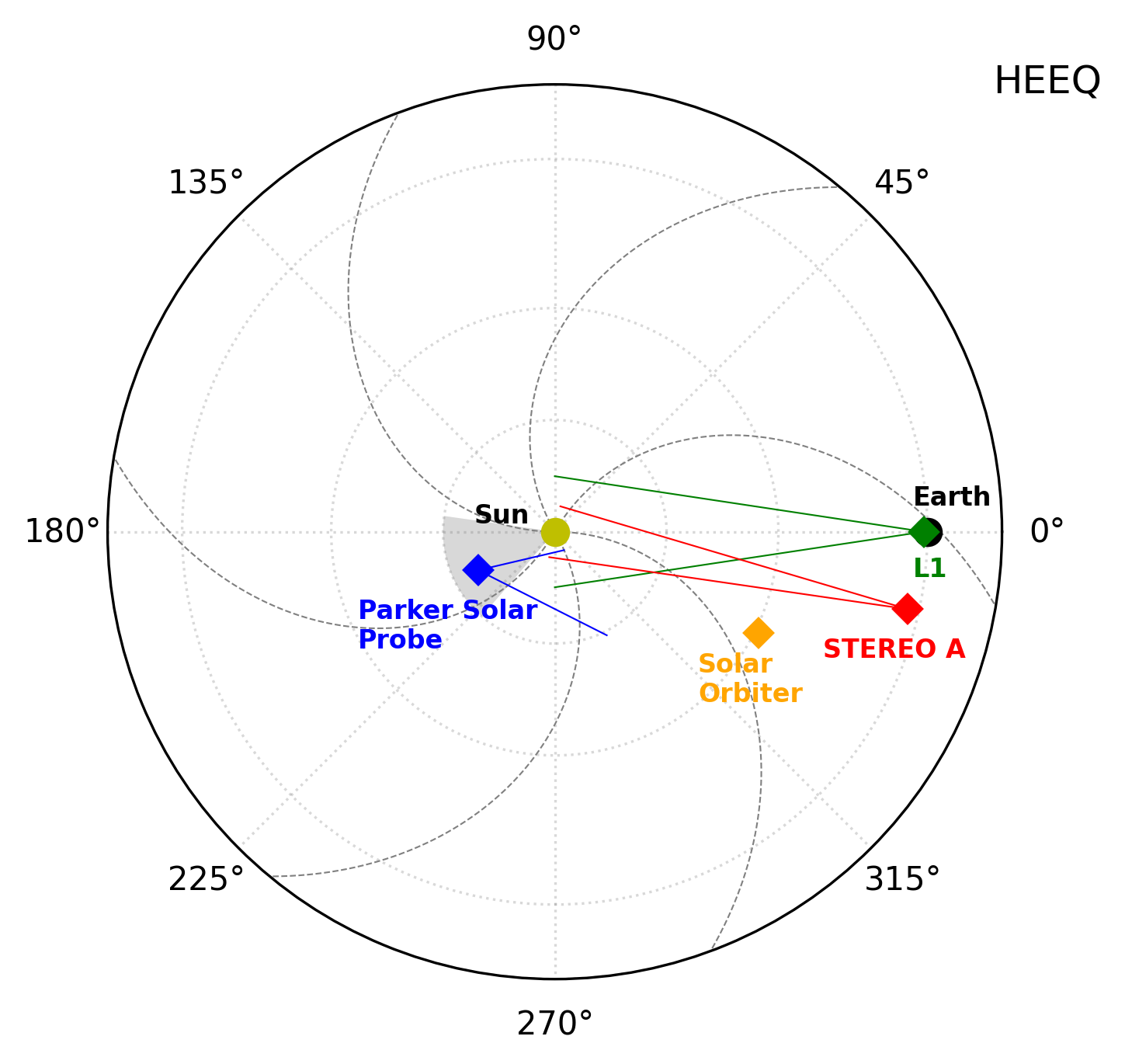}
    \caption{Spacecraft locations in HEEQ coordinate system on 2023 March 13 at 03:45 UT. The shaded area represents the eruption sector associated to the primary propagation direction.}
    \label{fig:position}
\end{figure}

The CME-driven shock was observed by different spacecraft, including Parker Solar Probe (PSP), located on the far side of the Sun as seen from Earth (Figure \ref{fig:position}). PSP observed a very intense SEP event, so the source region of the emission is likely located on the opposite side of the Earth (see the shaded sector in Figure \ref{fig:position}). Radio data from PSP detected a series of fast-drifting type III bursts, followed by a strong, slower-drifting type II radio burst. This emission is a sign of shock formation and suggests that the main eruption at the Sun occurred at 03:13 UT \citep{jebaraj_2024,dresing_2025}. Due to the absence of direct solar observations from the far side, \citet{dresing_2025} identify two candidate active regions, one in the northern and the other in the southern hemisphere, as the source of the widespread event. 

Thanks to the possibility of observing the CME-driven shock on 2023 March 13 from different spacecraft, we are actually able to reconstruct the shock front in 3D.

\subsection{Remote-sensing observations}
To characterize the shock front in the solar corona, we focused on the analysis of the images acquired by the COR2 coronagraph of STEREO-A \citep{kaiser_2008}, which has a field of view (FoV) extending between 2--15 R$_\odot$, and the LASCO/C3 \citep{brueckner_1995} coronograph of SoHO, with a FoV up to 32 R$_\odot$.
Additionally, the shock expands through the FoVs of the PSP/WISPR \citep{vourlidas_2016} and STEREO-A/HI-1 and HI-2 heliospheric imagers (not shown here).

\begin{figure}
        \centering
        \includegraphics[width=1.0\linewidth]{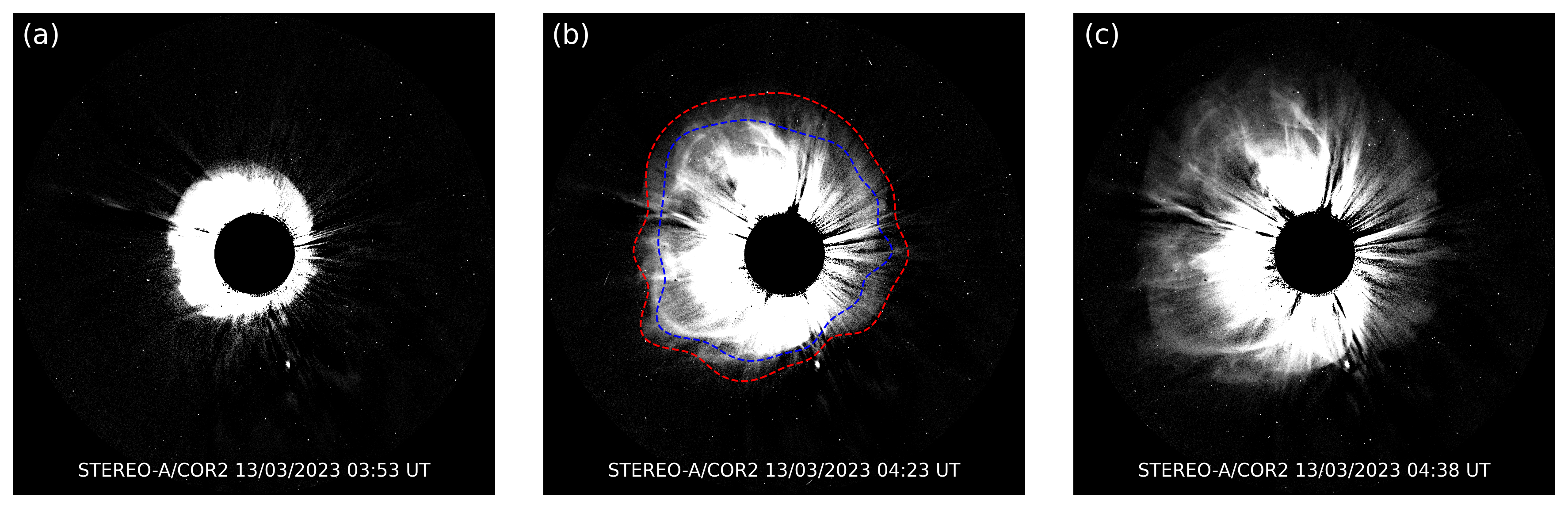}
        \caption{Total brightness of base difference images taken on 2023 March 13 at 03:53 UT (a), 04:23 UT (b) and 04:38 (c) from STEREO-A/COR2. In panel (b) dashed lines refer to the shock front (red curve) and flux rope (blue line). The threshold intensity has been chosen with the minimum value equal to zero, in order to darken the background and to highlight the presence of the shock in front of the CME.}
        \label{fig:cor2_shock_front}
    \end{figure}

COR2 FITS data are downloaded\footnote{\url{https://stereo.gsfc.nasa.gov/}} in the time interval 03:53:00-04:53:00 on 2023 March 13. We use total brightness images calculated from double-polarization data. The image data are acquired with a cadence of 15 minutes and have a spatial resolution of 14.7 arcsec, equivalent to a projected distance of about $10^4$ km. The data are calibrated using the standard SolarSoftWare (SSW) routine \texttt{secchi\_prep}, which performs corrections to the data, such as exposure-time normalization, vignetting function, and conversion of the intensity from digital number (DN) to mean solar brightness (MSB).
To highlight the appearance of the shock in front of the CME leading edge, we perform a base difference of the image data set by taking the COR2 image at 02:53:00 as the base image. This procedure is expected to remove the contribution of the F-corona and the brightness of background streamers. Figure \ref{fig:cor2_shock_front} shows the propagation of the shock in base difference images from STEREO-A/COR2 at three different times: 03:53 UT, 04:23 UT, and 04:38 UT, corresponding to panels (a), (b), and (c), respectively. The shock front appears as a faint contour outside of a bright envelope that represents the flux rope (see panel (b) in Figure \ref{fig:cor2_shock_front}). The CME-driven shock expands rapidly, covering in about one hour the entire FoV of COR2. The shock is not yet visible at 03:53 UT, and becomes clear 30 minutes later (see the dashed red line in panel (b) in Figure \ref{fig:cor2_shock_front}), detached from the apparent leading front of the CME (the dashed blue line in panel (b) in Figure \ref{fig:cor2_shock_front}). The event appears as a halo in both the FoV of STEREO-A/COR2 and SoHO/LASCO/C3 (e.g., see Figure \ref{fig:geometric_profile}), although the angular separation between the spacecraft is not very large (about 8.4 deg).

\subsection{In-situ data}
\begin{figure}[htpb]
    \centering
    \includegraphics[width=0.4\linewidth]{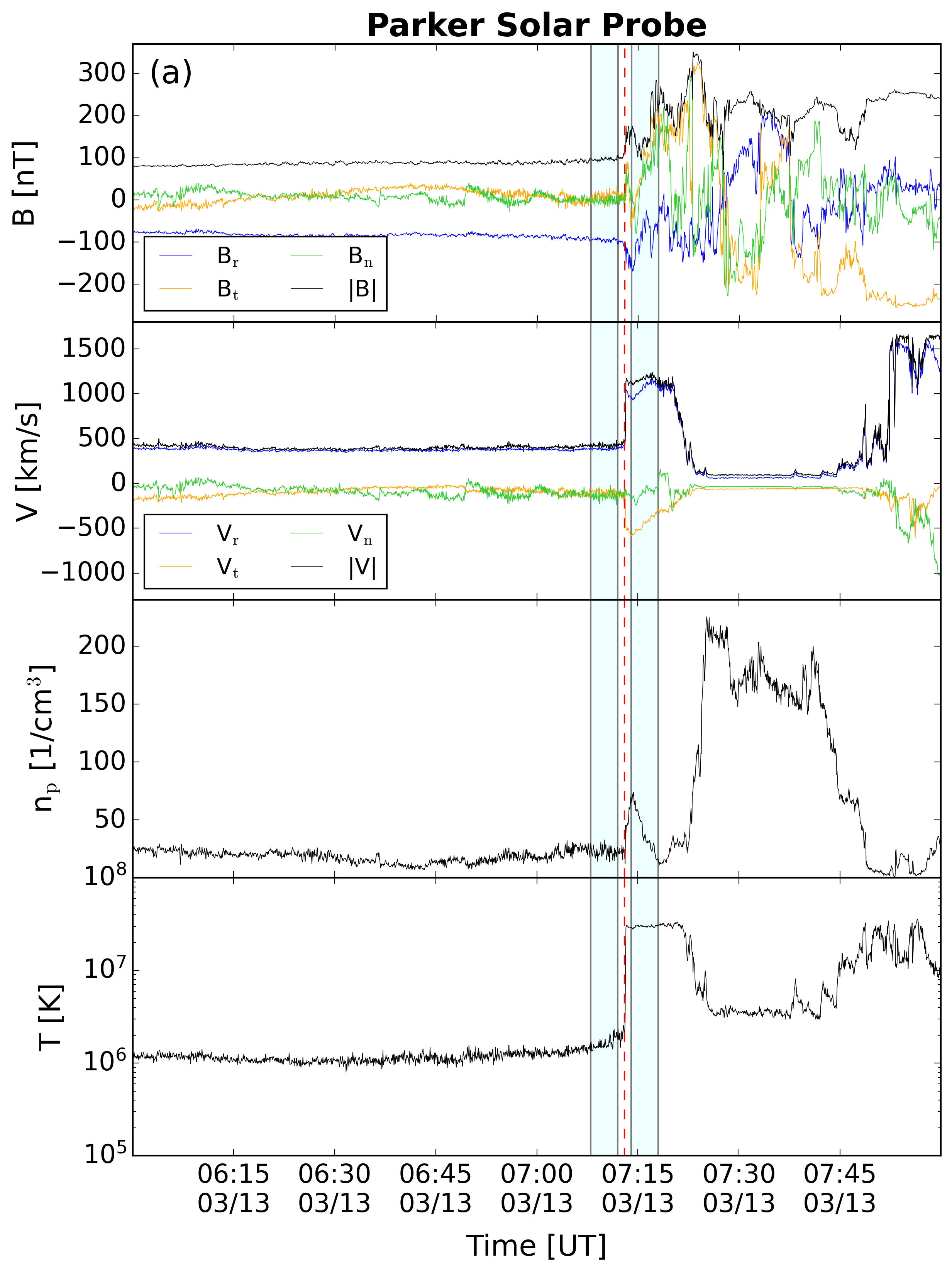}
    \includegraphics[width=0.4\linewidth]{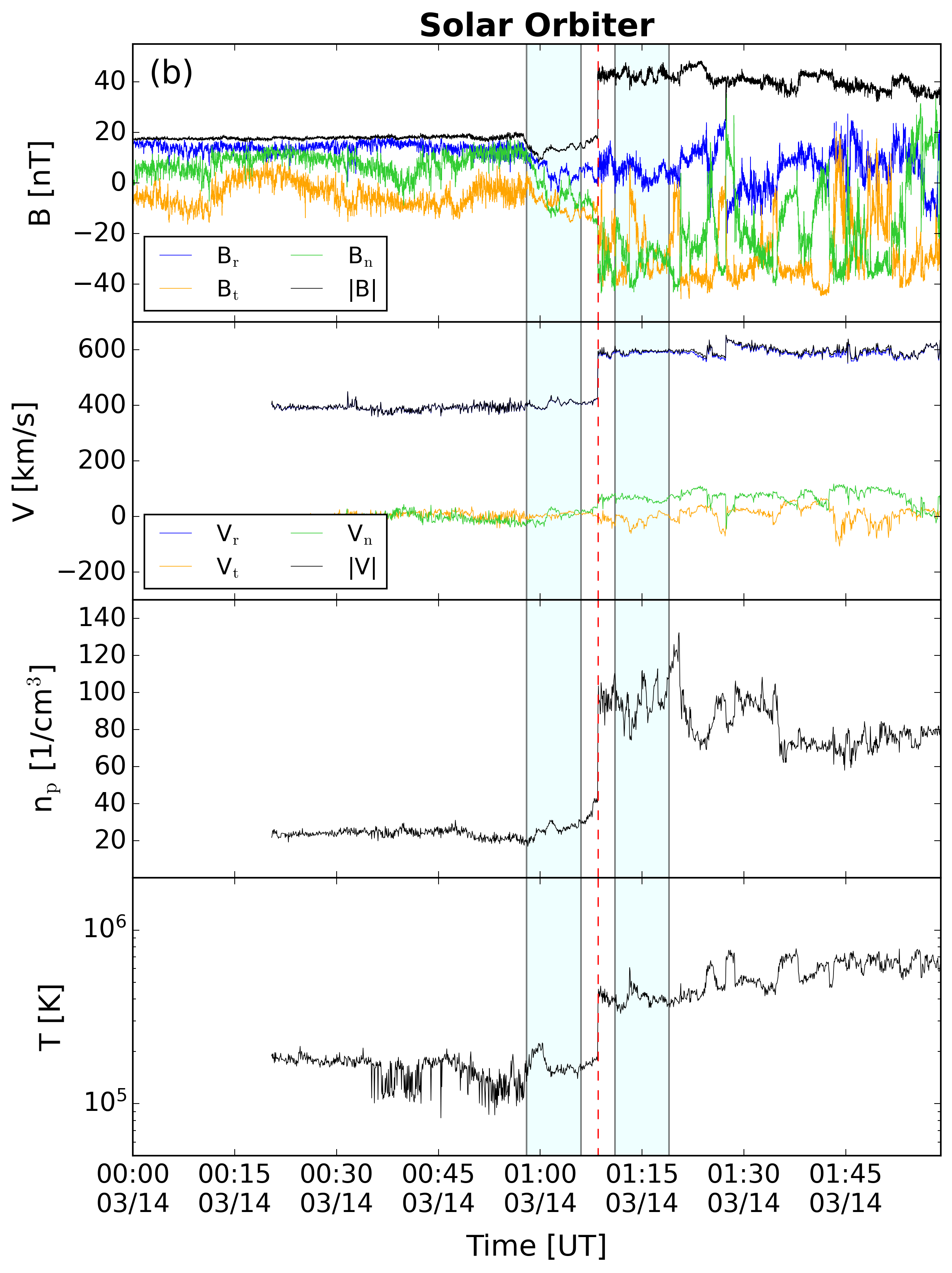}
    \includegraphics[width=0.4\linewidth]{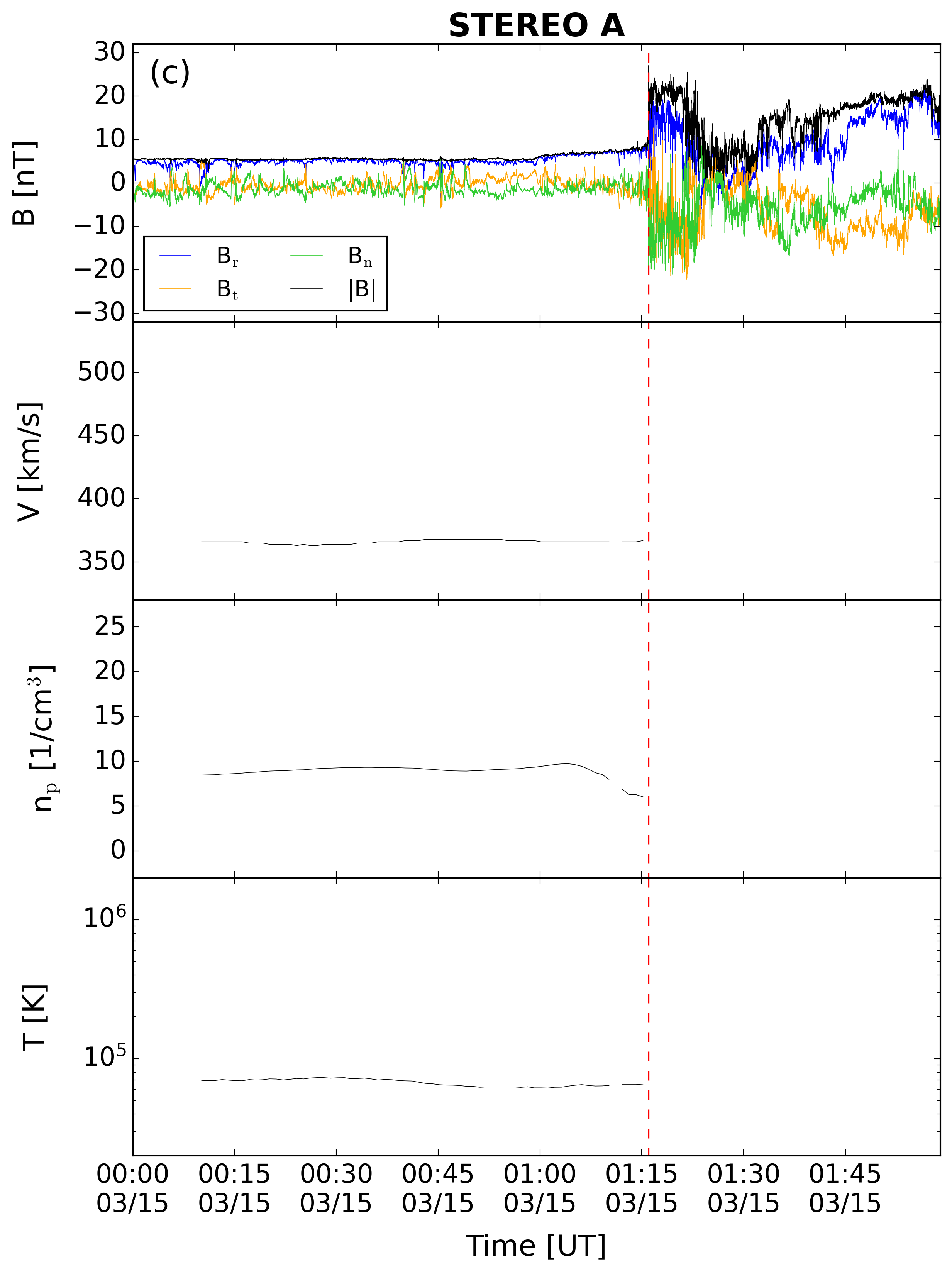}
    \includegraphics[width=0.4\linewidth]{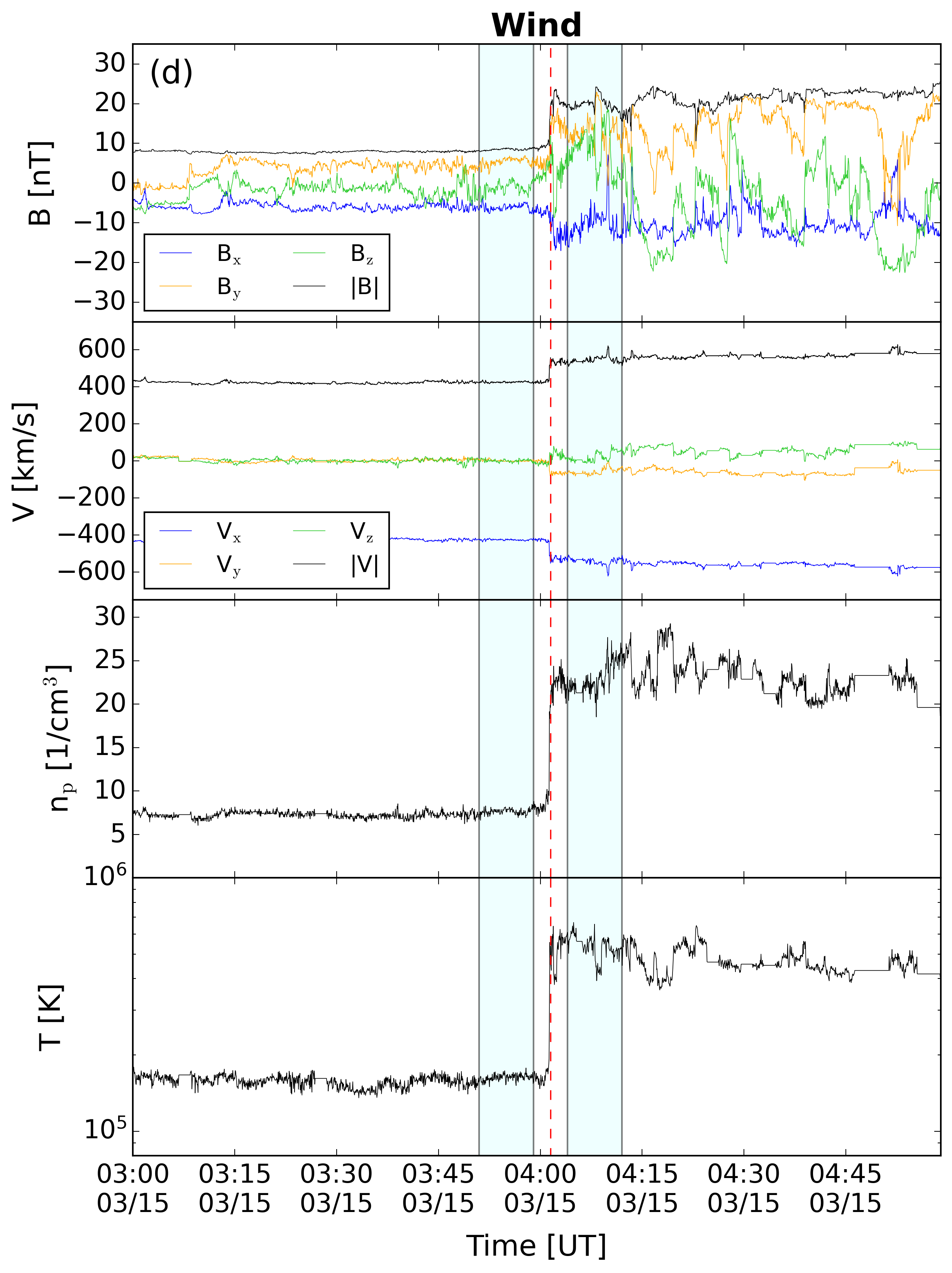}
    \caption{Magnetic field and plasma observations by PSP (a), SolO (b), STEREO-A (c), and Wind (d). From top to bottom, the panels show the magnetic field components and magnitude, the bulk velocity components and magnitude, the proton density and the plasma temperature. The vertical dashed lines indicate the time of the spacecraft shock crossings, PSP on 13 March at 07:13 UT, SolO on 14 March at 01:08 UT, STEREO-A on 15 March at 01:16 UT, and Wind on 15 March at 04:02 UT, respectively. The shaded areas mark the time interval in the upstream and downstream regions to calculate the shock parameters.}
    \label{fig:overview-shocks}
\end{figure}

As reported in Figure \ref{fig:position}, the spacecraft were largely spread in longitude. The shaded sector indicates the most probable region for hosting the eruption associated with this halo CME \citep{dresing_2025}. Thus, with such a wealth of in-situ measurements, it is possible to characterize this event from different shock crossings. For PSP we make use of 1 min resolution observations of the magnetic field, obtained from the fluxgate magnetometer, part of the FIELDS suite \citep{Bale_2016SSRv}, and of 4 s cadence solar wind measurements, provided by the Solar Probe Analyzers (SPAN) instrument, which is part of the Solar Wind Electrons Alphas and Protons investigation \citep[SWEAP; ][]{Kasper_2016SSRv}. For Solar Orbiter we analyze magnetic field time series from the magnetometer \citep[MAG; ][]{Horbury_2020} with a resolution of 0.125 s and 4 s plasma data from the Solar Wind Analyser \citep[SWA; ][]{Owen_2020} Proton and Alphas Sensor (PAS) (sampling 3D velocity distribution functions of protons and alpha particles). Magnetic field and solar wind plasma measurements on board STEREO-A are obtained from the Magnetic Field Experiment (MFE), part of the In situ Measurements of Particle and CME Transients \citep[IMPACT; ][]{Luhmann_2008} and from the Plasma and Suprathermal Ion Composition \citep[PLASTIC; ][]{Galvin_2008SSRv} instruments with a resolution of 0.125 s and 4 s, respectively. For Wind, 3 s magnetic field data are provided by the Magnetic Fields Investigation \citep[MFI; ][]{Lepping_1995SSRv} magnetometer, and 3 s plasma data from the Three-Dimensional Plasma Analyzer \citep[3DP; ][]{Lin_1995SSRv}. Overviews of the magnetic field, the bulk solar wind velocity, proton density, and temperature recorded by PSP, Solar Orbiter, STEREO A, and Wind are shown in Figure \ref{fig:overview-shocks}. The dashed lines refer to the shock crossing over the four spacecraft on 13 March at 07:13 UT, 14 March at 01:08 UT, 15 March at 01:16 UT and 15 March at 04:02 UT, for PSP, Solar Orbiter, STEREO-A, and Wind, respectively. The shaded areas indicate the time interval in the upstream and downstream regions of the shock used to calculate key parameters associated with the shock passage. We exclude the STEREO-A observations for the in-situ analysis due to the lack of plasma data in the downstream region of the shock.

The key parameters associated with the shock crossing of the three spacecraft (PSP, Solar Orbiter, and Wind) are derived from the calculation of the normal to the shock direction using the mixed mode method \citep{2000ESASPPaschmann}, within intervals of 8 minutes in the upstream and downstream regions of the shock after removing 5 minutes around the shock transition through Solar Orbiter and Wind. For PSP these quantities are calculated using intervals of 5 minutes upstream and downstream of the shock passage, excluding 2 minutes around the crossing (see shaded areas in Figure \ref{fig:overview-shocks}). These time intervals are chosen to exclude the density increase to about 200 cm$^{-3}$ associated to an almost zero solar wind velocity. This limitation is due to the field of view of the SPAN-I instrument. The gas compression ratio is defined as the ratio between the downstream ($n_d$) and the upstream ($n_u$) number densities ($X=n_d/n_u$), the magnetic compression ratio as the ratio between the downstream and upstream magnetic field ($r_{\text{B}} =\text{B}_d/\text{B}_u$), the plasma beta as the ratio between kinetic pressure and magnetic pressure ($\beta = p/p_{\text{B}}$), the Alfv\'enic and magnetosonic Mach numbers as the ratio between the shock speed in the upstream reference frame and the Alfv\'en and fast magnetosonic speeds, respectively (M$_{\text{A}} = v_{sh}/v_{\text{A}}$ and M$_{\text{ms}} = v_{sh}/v_{\text{ms}}$). The key parameters are calculated using a running window inside the chosen intervals. The values and the errors are obtained by taking the mean (or the median when the distribution is widely spread) and the standard deviation of the different quantities. The obtained shock parameters with their relative errors are given in Table \ref{tab:parameters}. The columns report for each spacecraft the heliocentric distance from the Sun, the date and time of the shock crossing, the shock-normal angle ($\theta_{\text{Bn}}$), the compression ratio ($X$), the magnetic compression ratio ($r_{\text{B}}$), the plasma beta ($\beta$), the Alfv\'enic (M$_{\text{A}}$) and the magnetosonic (M$_{\text{ms}}$) Mach numbers.
For PSP observations, while the determination of the normal to the shock vector is in very good agreement with \citet{jebaraj_2024,dresing_2025} (namely, it lies almost along the radial direction), the values of the gas compression ratio and of the Alfv\'enic Mach number have been found to be lower than those found in previous works due to an underestimation of the gas density upstream of the shock from SPAN-i. For an alternative estimation of the density, we check the presence of Langmuir waves in the waveforms acquired by the Time Domain Sampler (TDS) onboard PSP (see \citet{Bale_2016SSRv}) and determine the electron density from the frequency at which the electric field power spectral density shows a clear maximum, if present. While we observe a clear signature of Langmuir waves upstream, as in \citet{jebaraj_2024}, we cannot infer any value downstream because of the lack of such Langmuir wave signatures. If we assign to the downstream density its value at the shock from the TDS, we obtain a gas compression ratio of about 2.54, closer to the result by \citet{jebaraj_2024}. Thus, in both the density estimations an approximated value can actually be obtained.

We also analyze the magnetic field fluctuation power during each shock crossing (see the Appendix for additional details). The level of magnetic field fluctuations just upstream and downstream of the front can actually vary from satellite to satellite, implying that spacecraft are possibly connected to different shock regions. For example, during the PSP and STEREO-A crossings the power spectral densities (PSDs) are spread over a broad number of time scales, from large (about $100$ s) towards short time scales in the upstream (these two crossings have also very similar values of the magnetic compression, $r_B\sim 1.78$ at PSP and $r_B\sim 1.55$ at STEREO-A). During Solar Orbiter and Wind crossings the PSD is mostly high at large time scales. This suggests that the variation of the shock parameters at different spacecraft locations can indeed be influenced by the presence of different levels of magnetic turbulence. 

On the other hand, the shock parameters derived in this work are in good agreement with the quantities estimated in the SERPENTINE Solar cycle 25 SEP Events Catalog \citep{trotta2024} and in the CfA Interplanetary Shock Database\footnote{\url{https://lweb.cfa.harvard.edu/shocks/wind.html}}.

\begin{table}[htpb]
    \centering
    \caption{Parameters of the shock observed from PSP, SolO, and Wind. Columns: satellite, heliocentric distance from the Sun, date and time of the shock, shock-normal angle ($\theta_{\text{Bn}}$), compression ratio ($X$), magnetic compression ratio (r$_{\text{B}}$), beta of the plasma ($\beta$), Alfv\'enic (M$_{\text{A}}$) and magnetosonic (M$_{\text{ms}}$) Mach numbers.}
    \label{tab:parameters}
    \vspace{0.2cm}
    \begin{tabular}{c|c|c|c|c|c|c|c|c}
    \hline
    \hline
     S/C & R [AU] & DATETIME & $\theta_{\text{Bn}}$ [°] & $X$ & r$_{\text{B}}$ & $\beta$ & M$_{\text{A}}$ & M$_{\text{ms}}$ \\
    \hline
      PSP & 0.23 & 13/03/2023 07:13 UT & 19.6 $\pm$ 9.1 & 1.39 $\pm$ 0.70 & 1.78 $\pm$ 0.55 & 0.28 $\pm$ 0.02 & 4.68 $\pm$ 3.45 & 4.23 $\pm$ 3.13 \\
      SolO & 0.60 & 14/03/2023 01:08 UT & 57.6 $\pm$ 24.1 & 3.60 $\pm$ 0.43 & 3.38 $\pm$ 0.40 & 1.97 $\pm$ 0.69 & 4.81 $\pm$ 0.58 & 2.99 $\pm$ 0.19 \\
      Wind & 0.99 & 15/03/2023 04:02 UT & 71.5 $\pm$ 10.3 & 2.98 $\pm$ 0.17 & 2.31 $\pm$ 0.15 & 1.14 $\pm$ 0.03 & 3.04 $\pm$ 0.20 & 2.17 $\pm$ 0.14 \\
    \hline
    \end{tabular}
\end{table}

\section{Physical model} 
\label{sec:model}
To have a deep understanding of the physics and of the dynamics of the shock, we want to determine how the shock parameters change at different distances from the Sun. Therefore, we complement in-situ data with remote sensing data. However, inferring the compression ratio from coronagraphic images is not straightforward but requires a valid modeling of the density profile on the shock surface and a comparison of a simulated brightness with the observed one. This can be done via raytracing simulations of the Thomson scattered emission \citep{thernisien_2006, Nistico2020}, as we will explain in the next paragraphs.

\subsection{Shock model}
    \begin{figure} 
        \centering
        \includegraphics[width=0.9\linewidth]{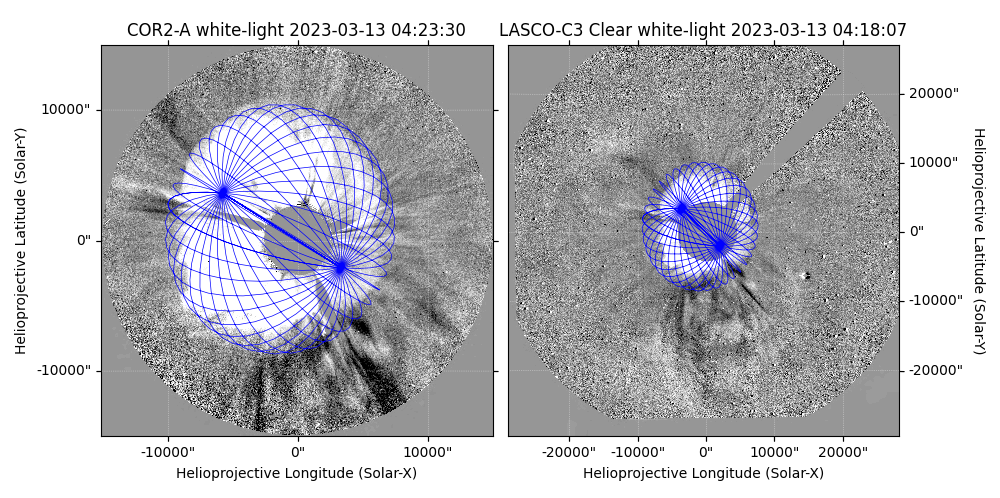}
        \caption{Quasi simultaneous observations of the CME on 2023 March 13 by STEREO-A/COR2 at 04:23 UT and SoHO/C3 at 04:18 UT. Blue curves represent the shock front modeled with the ellipsoid model. The images have been adapted from the ``CGS in Python" tool, available at \url{https://github.com/johan12345/gcs\_python}.}
        \label{fig:geometric_profile}
    \end{figure}

To build the geometric model of the shock, we adopt the Heliocentric-Earth-EQuatorial (HEEQ) reference frame, with the frame axis $X_{HEEQ}$ directed along the Sun-Earth line, $Z_{HEEQ}$ along the north direction of the Sun's rotation axis, and $Y_{HEEQ}$ perpendicular to them.     
Based on the work of \cite{kwon-vourlidas_2017}, we start by defining the expanding ellipsoid with the following parametric equations in the canonical form:

\begin{align}
    \left\{ \begin{array}{l l}
    x =&  a \cos u \cos v \\
    y =&  b \sin u \cos v \\
    z =&  c \sin v
    \end{array}
    \right.
    \label{canonical_ell}
\end{align}

with $x$, $y$, $z$ the coordinates of the points in the HEEQ reference frame. The quantities $u$ and $v$ define a 2D angular grid ranging in the intervals $0^\circ \le u \le 360^\circ$ and $-90^\circ \le v \le 90^\circ$. Initially, the center of the ellipsoid falls in the origin of the HEEQ system, and the ellipsoid semi-axes of length $a, b, c$ are aligned with the coordinate system axes (or in other words, the ellipsoid-fixed reference frame coincides with the HEEQ one). However, to properly fit the shock in the FoV of coronagraphs (see Figure \ref{fig:cor2_shock_front}), we should take into account the position of the center of the observed shock, which would be located high in the corona (ideally above the source region of the CME), and the proper orientation of the ellipsoid axes in the 3D space. To do that, we need to translate and rotate the ellipsoid-fixed frame with respect to the HEEQ coordinate system. In general, the 3D geometry of the ellipsoid would depend on nine free parameters (six for the ellipsoid axes and three for the position of its center). To simplify the reconstruction and to decrease the degrees of freedom, we force the  semi-axis $a$ to be directed outward from the chosen ellipsoid center along the solar radial direction, the semi-axis $c$ to lie in the plane containing the Sun's rotation and the $a$ axes, and $b$ is consequently determined to be perpendicular to both $a$ and $c$. Finally, given these constraints, the model depends on six parameters: the length of the axes $a$, $b$, $c$, and the position of the ellipsoid center defined in spherical coordinates by the triplet $(r, \theta, \phi)$. Therefore, starting from the canonical parametric equations in Eq. \eqref{canonical_ell}, the distribution of the points that represent the ellipsoid surface in the HEEQ coordinate system is defined by the following transformations:
\begin{itemize}

\item first, we perform a translation of the points obtained from Eq. \eqref{canonical_ell} along the semi-axis $a$ (that coincides with $X_{HEEQ}$) of a quantity $r_0$,
\begin{align}
    \left\{\begin{array}{l l}
    x' = & x + r_0 \\
    y' = & y \\
    z' = & z
    \end{array}
    \right.
\end{align}
\item Then, we perform two rotations of the points $(x',y',z')$: first in latitude around the $Y_{HEEQ}$ axis by an angle $\theta$, then in longitude around the $Z_{HEEQ}$ by an angle $\phi$.
\end{itemize}

The set of the transformed points 
%with coordinates
%$\mathbf{r}_{sh}=(x''', y''', z''')$%
will finally define the ellipsoid oriented in 3D space. By projecting the 3D model into the FoVs of the coronagraphic images,
%\sout{Using geometric triangulation with two or more different fields of view relevant to the different coronagraphic images analyzed,} 
we determine the free parameters of the model, i.e., the values of the axes $a$, $b$, $c$, the distance $r_0$ and the heliographic coordinates $(\theta,\phi)$ of the ellipsoid center. Figure \ref{fig:geometric_profile} shows the shock model superimposed on a pair of quasi-simultaneous base difference images from STEREO-A/COR2 (left panel) and SoHO/LASCO/C3 (right panel). The obtained free parameters of the ellipsoid model are given in Table \ref{tab:param-ellips}. We manually made several trials by changing the parameter values until we got good visual agreement between the model and the visible boundary of the shock in both FoVs of the instruments. Since this method consistently provides a reliable global shape of the shock with zero rotational angles of the ellipsoid axes at all time steps, the roll angle is therefore omitted from Table \ref{tab:param-ellips}.
%The method is not intended to provide an accurate reconstruction of the shock front, as this is also affected locally by irregularities (and we also exclude rotation angles of the ellipsoid axes, e.g. roll, pitch, and yaw angles), but to provide us with a reliable global shape of the shock envelope. 

\begin{table}
    \centering
    \caption{ Values of the axes $a$, $b$, $c$, the distance $r_0$ and the heliographic coordinates $(\theta,\phi)$ of the ellipsoid center.}
    \label{tab:param-ellips}
    \vspace{0.2cm}
    \begin{tabular}{c|c|c|c|c|c|c}
    \hline
    \hline
     TIME & $a$ [R$_\odot$] & $b$ [R$_\odot$] & $c$ [R$_\odot$] & $r_0$ [R$_\odot$] & $\theta$ & $\phi$ \\
    \hline
      03:53 UT & 5.50 & 5.00 & 5.50 & 2.00 & 23.88° & 199.15° \\
      04:23 UT & 9.80 & 8.70 & 9.80 & 3.00 & 23.88° & 199.15° \\
      04:38 UT & 12.20 & 10.80 & 12.20 & 4.00 & 23.88° & 199.15° \\
      04:53 UT & 14.00 & 12.70 & 14.00 & 4.50 & 23.88° & 199.15° \\
    \hline
    \end{tabular}
\end{table}

\subsubsection{The density model at the shock}\label{subsec:density}
\begin{figure}[htpb]
        \centering
        \includegraphics[width=0.5\linewidth]{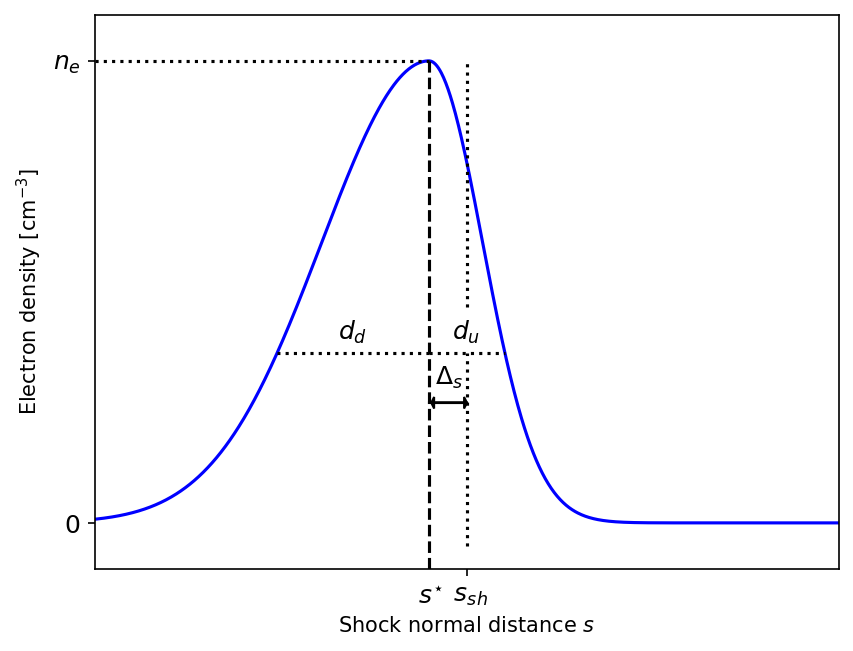}
        \caption{Sketch of the shock density profile as a function of the distance $s$ along the shock normal. The upstream and downstream widths of the profiles $(d_u,d_d)$, as well as the position of the shock front $s^\star$ are shown. We also indicate the position of the shock front $s_{sh}$ determined from the geometric modeling, and the shift $\Delta_s$, which is used to correct the location of the front based on the analysis of the brightness profiles (see text for further details).} 
        \label{fig:shock_profile}
    \end{figure}

Provided the geometry, a density profile on the surface is superimposed to simulate the Thomson-scattered emission of the shock. Following \citet{kwon-vourlidas_2018}, the profile has a double Gaussian-like shape with respect to the distance $s$, measured along the normal to the surface, namely (see Figure \ref{fig:shock_profile})

    \begin{equation}
        \centering
        n_{\rm sh}(s) = 
        \begin{cases}
            $ $ n_e \exp{\left[-\dfrac{(s - s^\star)^2}{d_d^2}\right]} & \text{for $s < s^\star$}\\
            $ $ n_e \exp{\left[-\dfrac{(s - s^\star)^2}{d_u^2}\right]} & \text{for $s > s^\star$}\\
        \end{cases}  
        \label{eqn:nsh}
    \end{equation}
    
with $n_e$ the density excess at the peak profile, $d_d$ e $d_u$ the widths of the Gaussian, $s^\star = s_{\rm sh} - \Delta_s$ the peak position, where $s_{\rm sh}$ is the position of the shock front determined by the geometric model and $\Delta_s$ is a quantity used to correct the density peak position in order to obtain the best-fit of the measured brightness profile with those from the raytrace simulations (see Sec. \ref{sec:compression_ratio}).

\subsection{Raytracing simulation}
The faint emission that characterizes a shock front is determined by the integrated column density along the line of sight. Raytracing simulations of the Thomson scattered emission, namely, the emissivity integrated along the LOS, represent a powerful tool for comparing theoretical models with observations. We perform raytracing applied to a datacube of density by using the routines available in SolarSoftWare \citep[SSW; ][]{thernisien_2006, thernisien_2011}. We create synthetic images of STEREO-A/COR2 that are finally analysed and compared to the observations.

\begin{figure}[htpb]
    \centering
    \includegraphics[width=0.9\linewidth]{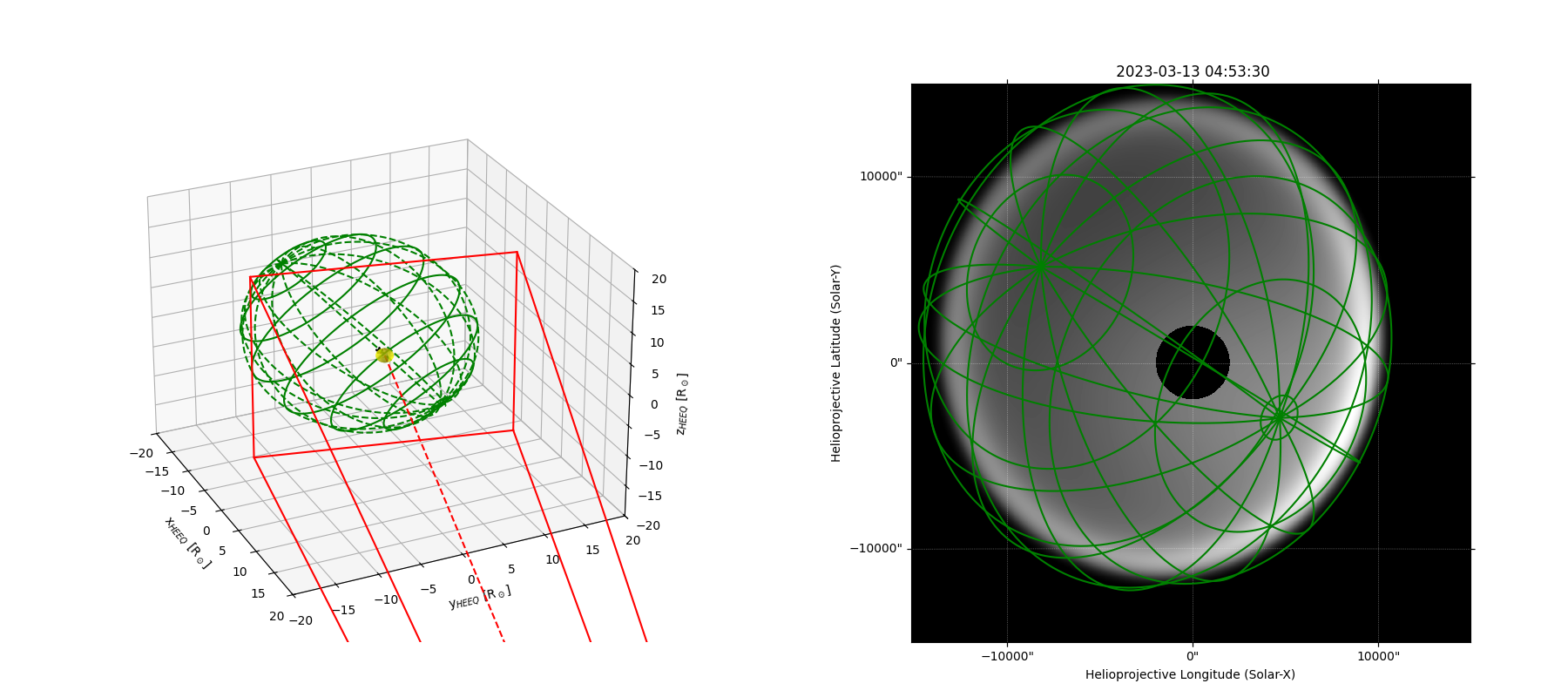}
    \caption{Left: 3D model of the ellipsoid representing the expanding shock in green. The partial frustum in red represents the FoV of STEREO-A/COR2, which is located at a distance of about 215 R$_\odot$ from the Sun. Right: Simulated brightness of the shock as seen from STEREO-A/COR2. The 3D model of the ellipsoid in green, which was used as an input to construct the density model of the shock, is superimposed on the simulated brightness image.}
    \label{fig:raytrace_example}
\end{figure}

Given the shock geometry and the set of parameters $(n_e, \Delta s, d_u, d_d) $ derived from the modeled electron density (Equation \ref{eqn:nsh}), we construct a density cube of size $L_x, L_y, L_z =$ 40 ${\rm R}_\odot$ and spatial step $\Delta = 0.1$ R$_\odot$ in the HEEQ coordinate system.  
%The sides of the density cubes are aligned with the HEEQ coordinate system (i.e., $L_{x}\parallel X_{HEEQ}$, etc). 
Therefore, the data cube is made of $401^3$ points. We construct a sample of about 3185 datacubes to be compared with a single observed brightness profile, each corresponding to a given density model defined by varying the values of the parameters. More precisely, $n_e\in [0.60\times 10^4,1.40\times 10^4]$ cm$^{-3}$ with steps of $0.2\times 10^4$ cm$^{-3}$, which fall in the range of values found by \citet{kwon-vourlidas_2018}, $d_u \in [0.05,0.35]$ R$_\odot$ with steps of 0.05 R$_\odot$, $d_d\in [0.1,0.7]$ R$_\odot$ with steps of 0.1 R$_\odot$, and $\Delta s \in [0, 3]$ R$_\odot$ with steps of 0.25 R$_\odot$ to better estimate the position of the shock peak. When varying the parameters, we force $d_u$ to be smaller than $d_d$. Then, we use the IDL routine {\tt raytracewcs.pro} that requires in input the temporal and spatial information stored in the FITS header of an image from STEREO-A/COR2. This allows us to take into account the location and the FoV of COR2. The Thomson scattered emission is integrated along a line of sight of 1024 points, ranging from -30 to 30 ${\rm R}_\odot$. The result of the raytracing is an image with a size of $2048\times2048$ pixels. 

An example is shown in Figure \ref{fig:raytrace_example}, where in the left panel a geometric representation of the expanding ellipsoid is depicted along with the FoV of STEREO-A/COR2; in the right panel an example of the 2D image obtained from the raytracing simulations is plotted together with the ellipsoid (shown in the left panel) superimposed and projected on the plane of the sky. Notice here that the brightest regions at the edge of the ellipsoid indicate the shock front, while the dark part should correspond to the region occupied by the flux rope that is not simulated with this model. The center of the ellipsoid is shifted with respect to the Sun position (the black circle).  Such simulated models will then be compared with observations from STEREO/COR2-A, as reported in Section \ref{sec:compression_ratio}.
 
\section{Data analysis and results}
\label{sec:results}

\subsection{Kinematics}
\begin{figure}
    \centering
    \includegraphics[width=0.8\linewidth]{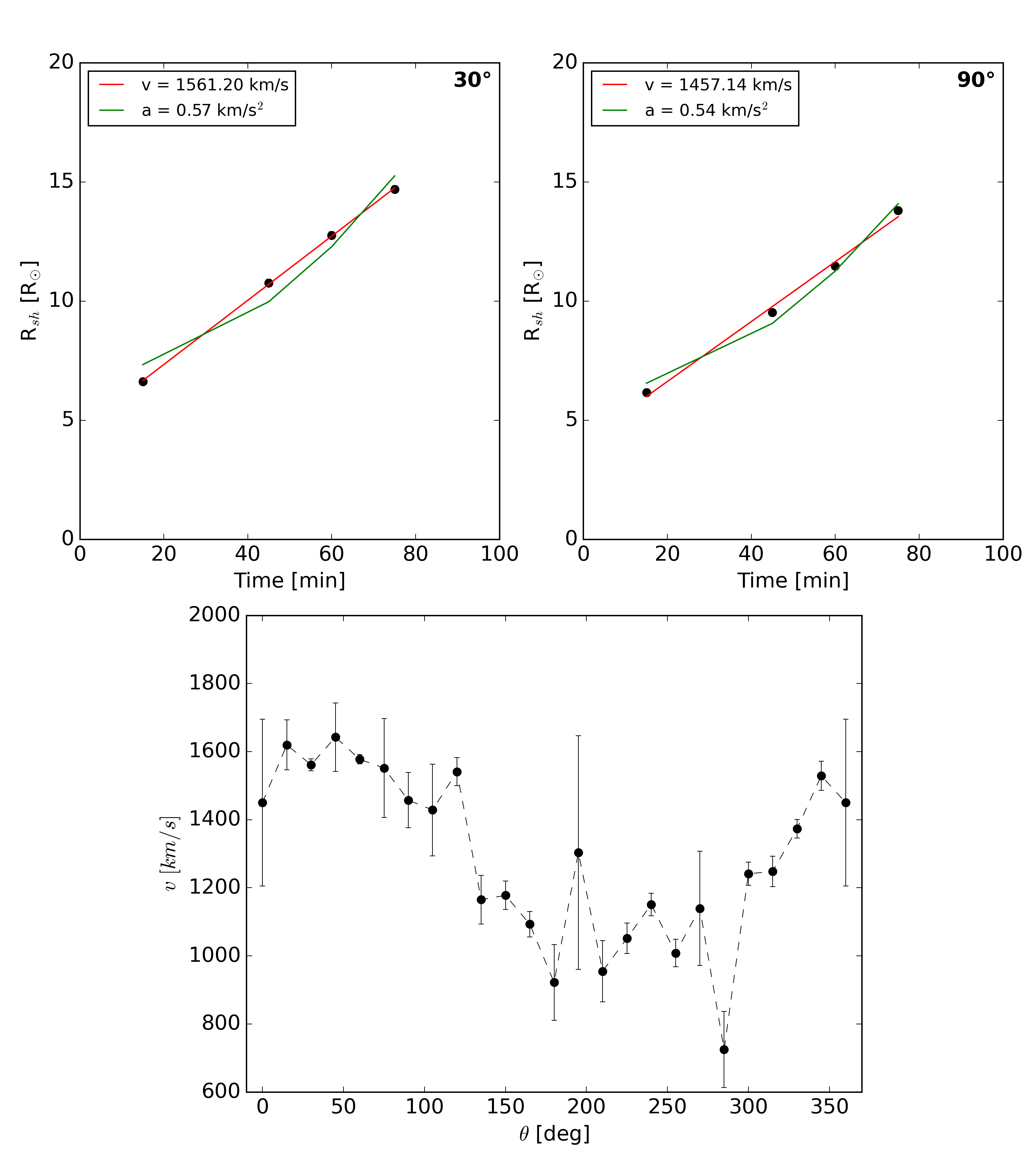}
    \caption{Upper panels: shock front position at four time steps for two different angles. Red and green lines represent the linear and quadratic fit. Data errors are within the marker size. Bottom panel: radial velocity derived for the 24 angles in the plane of sky.}
    \label{fig:kinematics}
\end{figure}

We investigate the kinematics of the halo CME in the time interval at which the shock front is visible in the STEREO-A/COR2 images. At each time, we took 24 radial cuts (relative to the center of the Sun and starting from the north pole) spaced by 15 degrees to estimate the propagation radial velocity. Upper panels in Figure \ref{fig:kinematics} show the linear and quadratic fit for two different angles (30° and 90°). Then we report the radial velocity obtained for the different angles in the bottom panel of Figure \ref{fig:kinematics}. It is evident that the fastest part of the CME is located on the north side (as seen from STEREO-A/COR2), namely small angles and close to $360^{\circ}$.

\begin{figure}
    \centering
    \begin{tabular}{c}
        \includegraphics[width=0.8\textwidth]{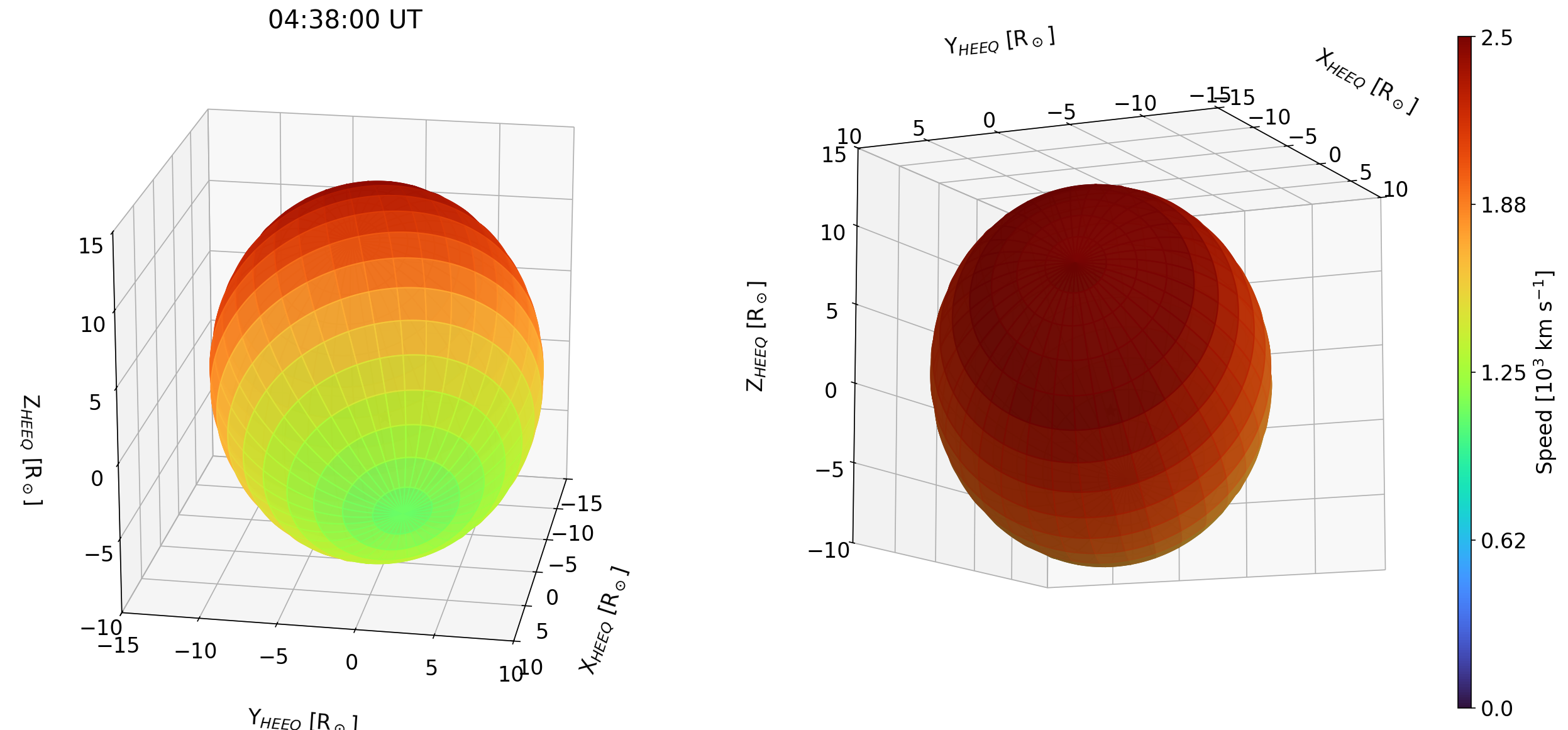} \\
        \includegraphics[width=0.8\textwidth]{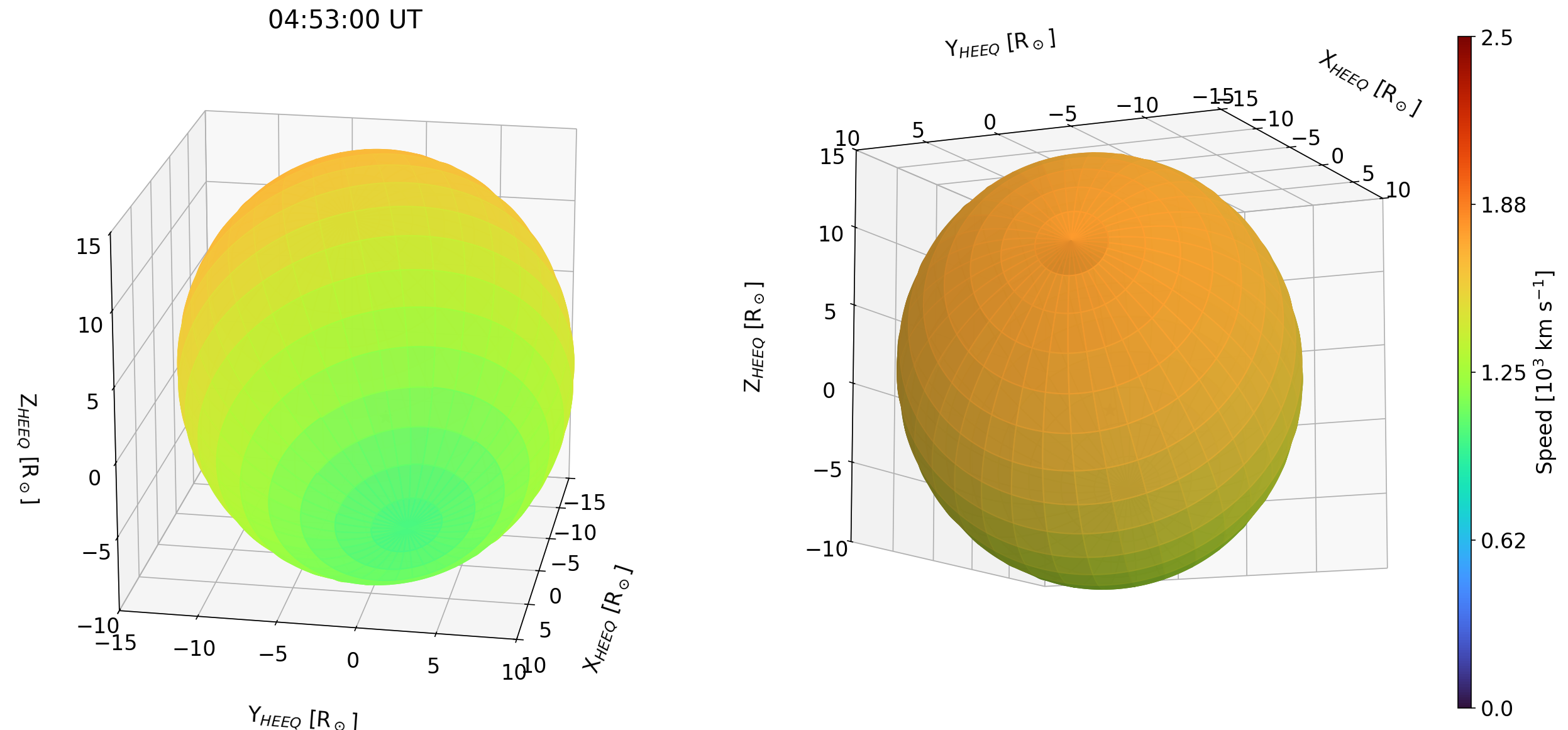}
    \end{tabular}
    \caption{Global speed map of the shock front at 04:38 (top panels) and 04:53 UT (bottom panels). The X$_{HEEQ}$ axis is directed along the Sun-Earth line, while Z$_{HEEQ}$ is parallel to the solar rotation axis. The maximum speed of the shock is mostly along the negative x direction.}
    \label{fig:speed_3d}
\end{figure}

On the other hand, a global characterization of the speed of the shock front can be based on the 3D geometric reconstruction. In Figure \ref{fig:speed_3d} we present the three-dimensional speed of the shock mapped on its front. Taking into account the position of each grid point of the ellipsoid at two consecutive time steps separated by $\Delta t$, that is $\mathbf{r}_{sh}(t)$ and $\mathbf{r}_{sh}(t-\Delta t)$, the velocity is computed as $\mathbf{v}_{sh}(t) = \left[\mathbf{r}_{sh}(t) - \mathbf{r}_{sh}(t- \Delta t)\right]/\Delta t$. Figure \ref{fig:speed_3d} shows the speed map on the shock surface at time 04:38 and 04:53 UT.
In both cases, the speed is computer over a time interval of 15 min.
In the top panels, the shock front has a maximum speed of $\sim$2500 km s$^{-1}$ at the nose. In the bottom panels, the shock front has lower speeds, ranging between 1800 and 1000 km s$^{-1}$, with the highest speeds at the shock nose. The 3D speed is coherent with those inferred in the FoV of STEREO-A/COR2 (hence affected by projection), and the position of the shock nose is compatible with the projected north sector of the shock as viewed by STEREO-A/COR2 where the highest projected speeds are estimated.

\subsection{Density excess and shock compression ratio}\label{sec:compression_ratio}

\begin{figure}
    \centering
    \includegraphics[width=0.8\linewidth]{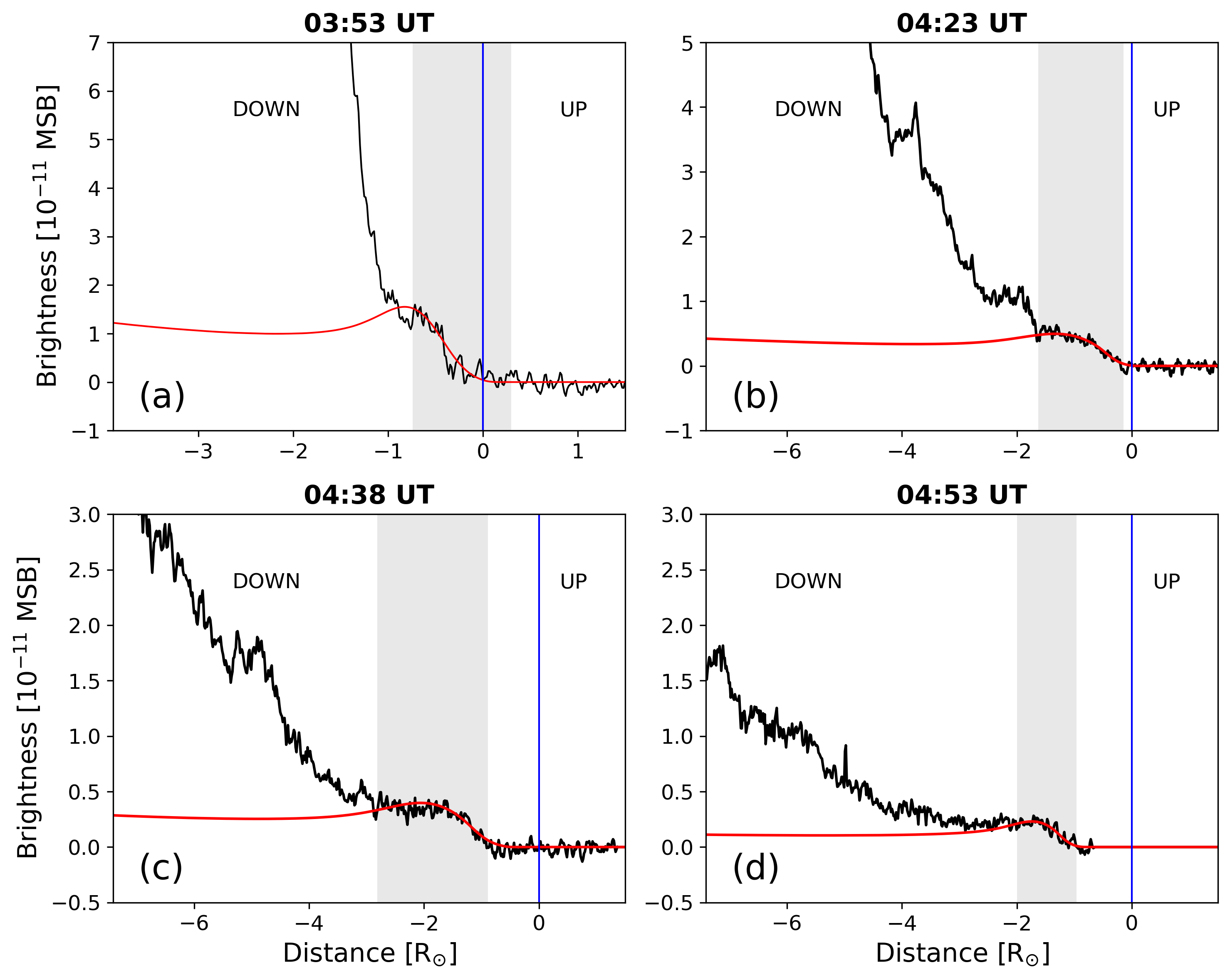}
    \caption{Comparison of the total brightness from the STEREO-A/COR2 base difference images at the four analyzed times (black line) and the excess brightness obtained from the best fit of our model (red line). The blue vertical line represents the shock front of the ellipsoid model. The grey shaded areas indicate the shock regions determined by visual inspection of the brightness data. The regions on the right of the shock front indicate the upstream (UP), while the regions on the left refer to the downstream (DOWN). The four panels correspond to the same cut at 30° at 03:53 UT (a), 04:23 UT (b), 04:38 (c) and 04:53 (d).}
    \label{fig:tB_profiles}
\end{figure}

\begin{table}
    \centering
    \caption{ Values of the parameters $(n_e, \Delta s, d_u, d_d) $ of our model that minimize the chi-squared for the observed total brightness profile around the shock region at 30° from the north direction for each time considered.}
    \label{tab:param-model}
    \vspace{0.2cm}
    \begin{tabular}{c|c|c|c|c}
    \hline
    \hline
     TIME & $n_e \times 10^4$ [cm$^{-3}$] & $\Delta s$ [R$_\odot$] & $d_u$ [R$_\odot$] & $d_d$ [R$_\odot$] \\
    \hline
      03:53 UT & 1.40$\pm$0.2 & 0.25$\pm$0.25 & 0.25$\pm$0.05 & 0.30$\pm$0.10 \\
      04:23 UT & 0.80$\pm$0.2 & 0.25$\pm$0.25 & 0.25$\pm$0.05 & 0.70$\pm$0.10 \\
      04:38 UT & 0.80$\pm$0.2 & 1.00$\pm$0.25 & 0.35$\pm$0.05 & 0.70$\pm$0.10 \\
      04:53 UT & 0.80$\pm$0.2 & 1.00$\pm$0.25 & 0.20$\pm$0.05 & 0.40$\pm$0.10 \\
    \hline
    \end{tabular}
\end{table}

\begin{figure}
    \centering
    \includegraphics[width=0.9\linewidth]{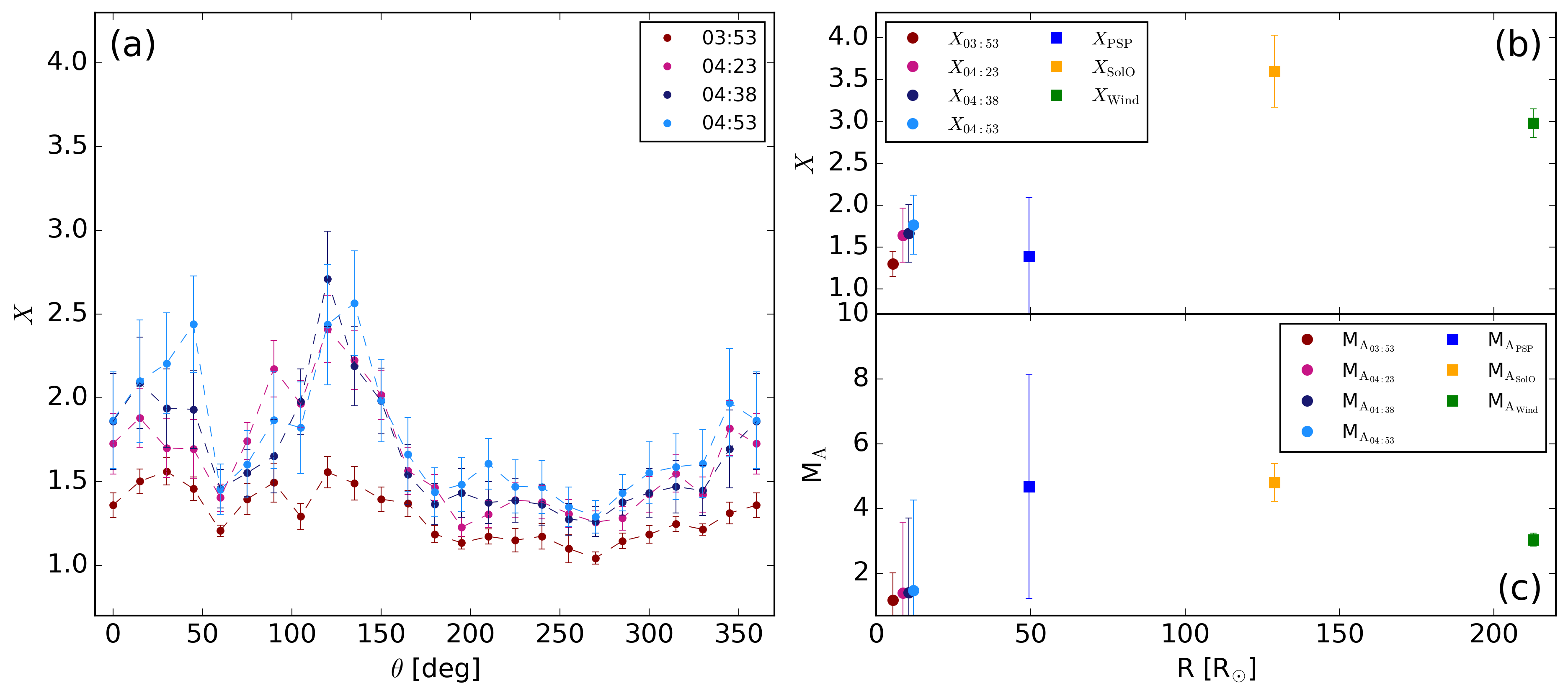}
    \caption{Compression ratio and Alfv\'enic Mach number evolution from remote sensing observations and in-situ measurements. Panel (a) shows the remote compression ratio derived for the 24 angles in the plane of the sky at four different times. Panels (b) and (c) show the radial evolution of the mean compression ratio and the Alfv\'enic Mach number at each time (circles), along with the values calculated in-situ at PSP, Solar Orbiter, and Wind (squares).}
    \label{fig:X_rtheta}
\end{figure}

\begin{figure}
    \centering
    \includegraphics[width=0.7\linewidth]{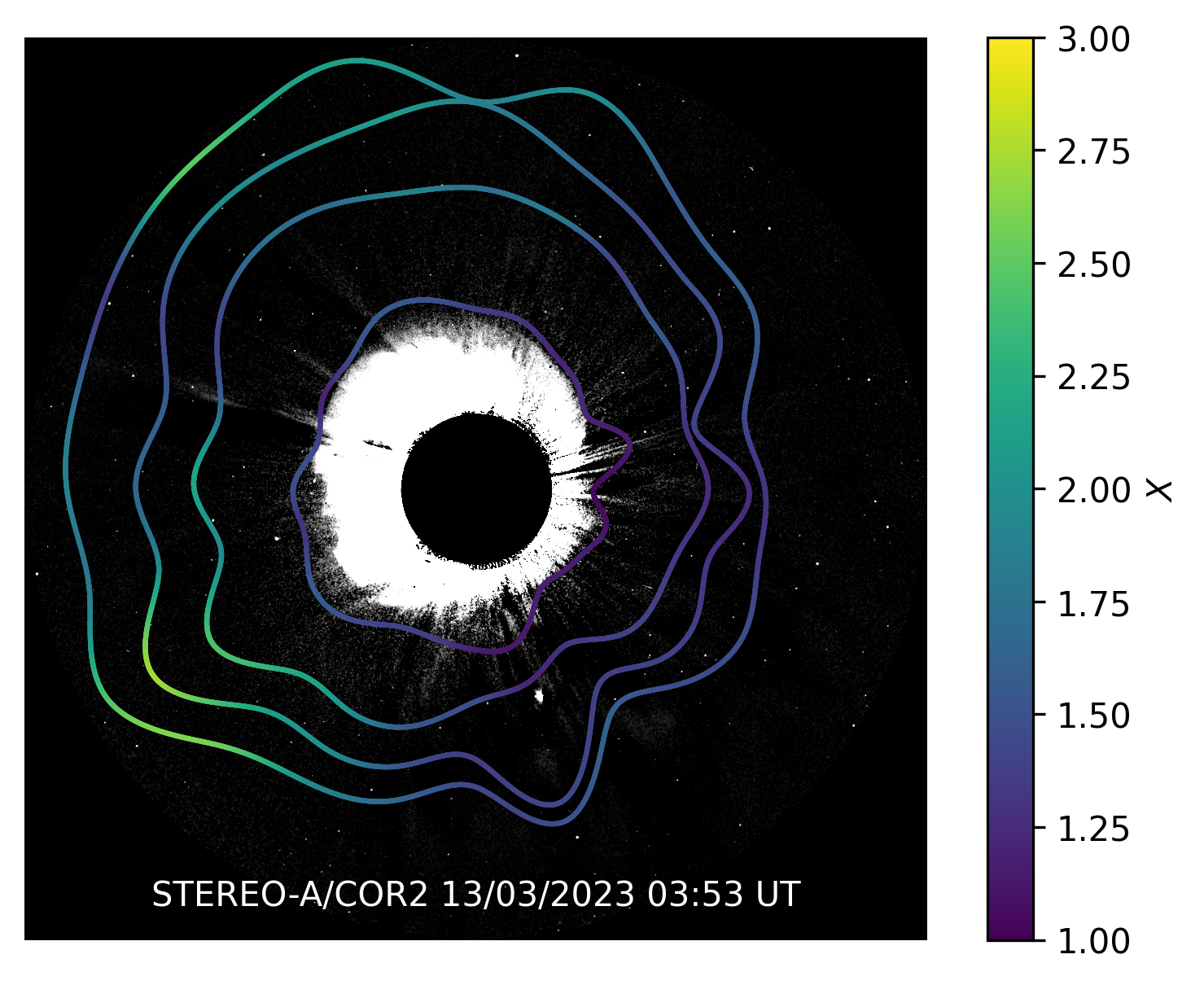}
    \caption{Total brightness image taken on 2023 March 13 at 03:53 UT from STEREO-A/COR2. Curved lines refer to the shock front positions at four different times. The color map refers to the compression ratio values.}
    \label{fig:X_map}
\end{figure}

We compare the total brightness derived from the base difference images with the synthetic ones obtained via raytracing simulations of the Thomson scattered emission. The comparison has been made for the four times at which the shock front is visible in the STEREO-A/COR2 images.
Since from the kinematic analysis we found that the region with higher velocity is in the north direction, we started from this position and proceeded counterclockwise to do the following analysis. In particular, at each time we took 24 boxes, normal to the shock surface modeled with the ellipsoid. The obtained parameters of our model that minimize the $\chi^2$ for the observed total brightness profile at 30° are given in Table \ref{tab:param-model}.

Figure \ref{fig:tB_profiles} shows the total brightness profiles (black line) obtained within the box, positioned at 30° and normal to the ellipsoid shock front, observed by STEREO-A/COR2 at 03:53 UT (a), 04:23 UT (b), 04:38 (c), and 04:53 (d). The blue vertical line represents the shock front position of the ellipsoid model, and the shaded areas mark the shock region, used for the density model fitting. On the right side of the shock front we localize the upstream region (UP), while on the left the downstream region (DOWN). Far downstream of the shock, the sharp increase in brightness is associated with the flux rope. The red lines indicate the best excess brightness obtained from our density model (notice that the flux rope is not actually included in the model). We perform a $\chi^2$ analysis to obtain the most suitable set of parameters of the double-Gaussian function $(n_e, \Delta s, d_u, d_d)$ from the observed brightness profile in the region of the shock; their values are reported in Table \ref{tab:param-model}. The $\chi^2$ fitting procedure has then been repeated for the 24 position angles.

The excess density $n_e$ at the shock peak is crucial for the determination of the compression ratio $X$. Therefore, we also need to evaluate the upstream (or background) density $n_u$. This can be obtained through the electron density inversion method applied to a polarized brightness image 
%at a given latitude angle in the plane of the sky
%by modeling it as an \textit{n}th-order polynomial function
\citep{vandeHulst_1950, hayes_2001}, which assumes an axisymmetric density distribution. As an electron density profile, we adopt the density model proposed by \cite{leblanc_1998}, namely

 \begin{equation}
      n_u(r)= a_2 \left(\dfrac{r}{{\rm R}_\odot} \right)^{-2} + a_4 \left(\dfrac{r}{{\rm R}_\odot} \right)^{-4} + a_6 \left(\dfrac{r}{{\rm R}_\odot} \right)^{-6},
      \label{eqn:leblanc}
  \end{equation}
  
which is a function of the radial distance from the Sun $r$ normalized to the solar radius ${\rm R}_\odot$, and $a_2$, $a_4$, and $a_6$ are the free parameters of the density model.
We use a polarized image from COR2 taken at 03:07:45 UT, just before the CME eruption, from a sequence of three images at linear polarization angles of 0$^\circ$, 120$^\circ$, 240$^\circ$ calibrated with the standard routine \texttt{secchi\_prep} in SSW and converted in MSB units. $pB$ images are a direct representation of the K-corona, since the interplanetary dust forming the F-corona is characterized by unpolarized emission. 

In particular, the electron density profiles upstream as a function of the radial distance from the Sun are obtained from the $pB$ for each considered azimuthal angle on the plane of the sky. They have been then fitted with the model in Equation (\ref{eqn:leblanc}). Thus, we get 24 upstream density profiles. To work with smoother quantities, we choose a $10^{\circ}$ window within which density profiles have been averaged. Once we obtain the upstream density, an estimation of the compression ratio through remote sensing observations is done by setting

  \begin{equation}
      X = \dfrac{\tilde{n_d}}{\tilde{n_u}} = \dfrac{n_e + \tilde{n_u}}{\tilde{n_u}},
      \label{eqn:X}
  \end{equation}
  
with $\tilde{n_u}$ and $\tilde{n_d}$ the values of the upstream and downstream densities found at the shock region in the COR2-A image.

The compression ratio values computed for all the 24 angles for each STEREO-A/COR2 image taken at the times indicated in Table \ref{tab:param-model} are shown in panel (a) in Figure \ref{fig:X_rtheta}, exhibiting slight variation with the latitude angle in the plane of the sky.
To determine the uncertainties in the compression ratio, we propagate the errors taking into account the grid resolution used to model the excess density (namely, 0.2$\times 10^4$ cm$^{-3}$) and the uncertainties in the coefficients $a_2$, $a_4$, and $a_6$ derived from fitting the polarized emission to the \cite{leblanc_1998} model. It is worth addressing that the choice of the grid resolution can affect the determination of the compression ratios, possibly leading to large uncertainties. Panel (b) shows the radial evolution of the compression ratio for both remote-sensing (circles) and in-situ (squares) data. For each of the remote observation times, we calculate the mean compression ratio over the entire shock front while the distance values $R$ are computed as the average distance of the shock front of the ellipsoid model from the center of the Sun. Although it is not possible to infer a robust trend in the estimates, we notice that the compression ratio tends to increase within 15 R$_\odot$. As the radial distance increases, the compression ratio rises up to values greater than 3, outlining that the shock remains strong. From in-situ observations, distances are taken from the spacecraft radial distances at the moment of the shock crossing.

In addition, the Alfv\'enic Mach number can also be determined from the compression ratio $X$, under the adiabatic assumption of $\gamma = 5/3$ \citep{bemporad_2011}. This parameter depends on the plasma $\beta$ and on the angle $\theta_{\text{Bn}}$. In the solar corona, the plasma is largely dominated by the magnetic field, so the plasma $\beta$, given by the ratio of kinetic pressure to magnetic pressure, can be considered negligible \citep{kwon_2013}. In this limiting case, we can estimate the Alfvénic Mach number for a perpendicular shock M$_{\text{A}\perp} = \sqrt{X(X+5)/[2(4-X)]}$ and for a parallel shock M$_{\text{A}\parallel} = \sqrt{X}$ \citep{vrvsnak_2002}. From the combination of these two relations, \cite{bemporad_2011} derived the Alfv\'enic Mach number for an oblique shock as 
\begin{equation}
    \text{M}_{\text{A}}^2 = (\text{M}_{\text{A}\perp} \sin\theta_{\text{Bn}})^2 + (\text{M}_{\text{A}\parallel} \cos\theta_{\text{Bn}})^2.
    \label{eqn:ma}
\end{equation}
For the evaluation of $\text{M}_{\text{A}}^2$ in corona, we use the mean compression ratio from remote data. Furthermore, we also choose the $\theta_{\text{Bn}}$ values as obtained from the in-situ analysis of the PSP shock crossing. This is actually an approximation, although from our shock model we found that in the corona the radial direction is almost aligned with the normal one (not shown). Thus, there is consistency between the quasi parallel shock found at PSP and the angle between the normal and the radial direction found in the corona, as deduced from the ellipsoid model. We obtain that M$_{\text{A}}$ is ranging between 1.18 and 1.48 in the time interval of STEREO-A/COR2 observations.  
Panel (c) in Figure \ref{fig:X_rtheta} shows the radial evolution of the Alfvénic Mach number for both remote-sensing (circles) and in-situ (squares) data. Regarding the Alfvénic Mach number as measured by PSP notice that it is affected by a large error. However, within the error bars the Alfvénic Mach number, for the $\beta$ and the $\theta_{\text{Bn}}$ values reported in Table \ref{tab:parameters} for PSP, is consistent with that compression ratio \cite[making use of Equation 7 in][]{vrvsnak_2002}.
We would like to stress that the methods used to compute the gas compression ratio and the Alfvénic Mach number from remote sensing and in-situ observations are different and comparison should be made with caution. Notice, for example, that the derivation of parameters from in-situ measurements does depend on how the spacecraft crosses the front and also on the level of turbulence around the shock, as shown in the Appendix.  

To better visualize how the compression ratio varies along the shock front at different times in the corona, Figure \ref{fig:X_map} superposes the total brightness image taken on 2023 March 13 03:53 UT from STEREO-A/COR2 with solid lines marking the shock front positions at four different times (see legend in Figure \ref{fig:X_rtheta}). Along the solid lines, values of the compression ratio estimated as a function of the longitude on the plane of the sky at each time are reported using a colorbar. From the values obtained at the 24 cuts, we make a cubic interpolation in order to have a continuous representation of the compression ratio. The color variation of the curves indicates how the compression ratio changes along the shock front. Indeed, it is possible to recognize regions of $X\sim 3.0$ (northern and southern) and zones of the shock front with very low compression (eastern and western). Notice that the regions where the shock seems to expand the most are the regions that show higher values of the compression ratio. In addition, according to the LASCO catalogue, the measurement position angle for this event is about $5^{\circ}$, which poses the fastest part of the leading edge in the north side, in agreement with our findings.

\section{Discussion and conclusions}
\label{sec:conclusions}
In this work, a peculiar CME event, which has been observed by several satellites, has been characterized in terms of its shock front evolution in the inner heliosphere. A very detailed analysis of the CME properties has been reported in \citet{dresing_2025}. By combining the remote-sensing observations from STEREO-A/COR2 and the in-situ observations both close to the Sun by PSP and at larger heliocentric distances up to 1 AU, it has been possible to track the shock front propagation in terms of the compression ratio and of the Alfv\'enic Mach number. Thanks to a technique that allows us to reconstruct the electron density on the shock front and to integrate along the line of sight, we determine from a minimization of the $\chi^2$ the excess density at the shock in the corona. Further, by analyzing the $pB$ at different radial cuts in the plane of the sky, we fit the upstream plasma density via a polynomial function proposed by \citet{leblanc_1998}, obtaining very good representations of the radial density profile far from the shock. Both the excess density evaluation at the shock and the upstream density are fundamental for the gas compression ratio determination along each radial cut and at different times. However, the values of the compression ratio in the corona can be dependent on the data cubes grid size and on the choice of the fixed grid for the evaluation of the density excess in the shock region; we plan to investigate such an aspect in a future work.
We have found a good agreement between the compression ratio determined in the corona and the one derived at the PSP radial distance.
%(once projected in the plane of the sky), 
However, this comparison needs a comment: the estimation of the gas compression ratio from PSP observations has been done using SPAN-i data, which actually represents a lower limit estimation, at least for the upstream side \citep{jebaraj_2024}.
Moreover, an additional source of variation of the shock parameters can be given by the way the satellite crosses the shock front; \citet{dresing_2025} discussed that PSP encountered the portion of the shock front associated with the flank side of the expanding CME ejecta. This can have an impact on the difference in the shock properties displayed in Table \ref{tab:parameters}.
In the Appendix we also show differences in the magnetic field fluctuations sampled by each satellite, addressing how the level of turbulent magnetic fluctuations can vary at different points of observation due to inhomogeneities in the interplanetary medium. This affects the shock parameter estimation: when the level of magnetic field power is large and spread over a broad range of time scales, as in PSP and STEREO-A (see Figure \ref{fig:wavelet} in the Appendix), a shock front distortion can be induced \citep{Trotta21} and high variability in the parameters is found along the shock front.
On the other hand, looking at the observations in corona, the values of the compression ratio along the shock front are almost homogeneous at a very early time (and typically $<2$), while they become and remain non-homogeneous at later times: in particular, $X$ reaches values up to $3$ in the northern and southern parts where the shock expands mostly (notice that from kinematics analysis the position of the nose of the shock has been found to be compatible with the projected north sector), while the compression ratio assumes very low values in the western and eastern sides. We have also found that at larger heliocentric distances, i.e., at $\sim 0.6$ AU and at Earth, the shock compression ratio remains $\ge 3$; probably the crossing spacecraft were connected with portions of the shock front characterized by high compression ratio values and during its evolution the shock wave did remain strong enough. Mapping the spatial and temporal evolution of the shock front in the corona is indeed fundamental also to better characterize whether in-situ satellites are connected to regions of higher or lower compression ratios and Mach numbers, and then to portions of the shock that are more or less efficient particle accelerators.

\begin{acknowledgments}
We acknowledge the project ‘Data-based predictions of solar energetic particle arrival to the Earth: ensuring space data and technology integrity from hazardous solar activity events’ (CUP H53D23011020001) ‘Finanziato dall’Unione europea – Next Generation EU’ PIANO NAZIONALE DI RIPRESA E RESILIENZA (PNRR) Missione 4 “Istruzione e Ricerca” - Componente C2 Investimento 1.1, ‘Fondo per il Programma Nazionale di Ricerca e Progetti di Rilevante Interesse Nazionale (PRIN)’ Settore PE09. GN, GZ, and SP acknowledge the Space It Up project funded by the Italian Space Agency, ASI, and the Ministry of University and Research, MUR, under contract n. 2024-5-E.0 - CUP n.I53D24000060005. SP and GZ acknowledge support by the Italian PRIN 2022, project 2022294WNB entitled “Heliospheric shocks and space weather: from multispacecraft observations to numerical modeling”. Finanziato da Next Generation EU, fondo del Piano Nazionale di Ripresa e Resilienza (PNRR) Missione 4 “Istruzione e Ricerca” - Componente C2 Investimento 1.1, ‘Fondo per il Programma Nazionale di Ricerca e Progetti di Rilevante Interesse Nazionale (PRIN)’. LSV received support by the Swedish Research Council (VR) Research Grant N. 2022-03352, by the International Space Science Institute (ISSI) in Bern, through ISSI International Team project \#23-591 (Evolution of Turbulence in the Expanding Solar Wind), and by the project “2022KL38BK– The ULtimate fate of TuRbulence from space to laboratory plAsmas (ULTRA)” (Master CUP B53D23004850006) by the Italian Ministry of University and Research, funded under the National Recovery and Resilience Plan (NRRP), Mission 4– Component C2– Investment 1.1, “Fondo per il Programma Nazionale di Ricerca e Progetti di Rilevante Interesse Nazionale (PRIN 2022)” (PE9) by the European Union– NextGenerationEU.
\end{acknowledgments}

\appendix \label{appendix}
\begin{figure}
    \centering
    \includegraphics[width=0.6\linewidth]{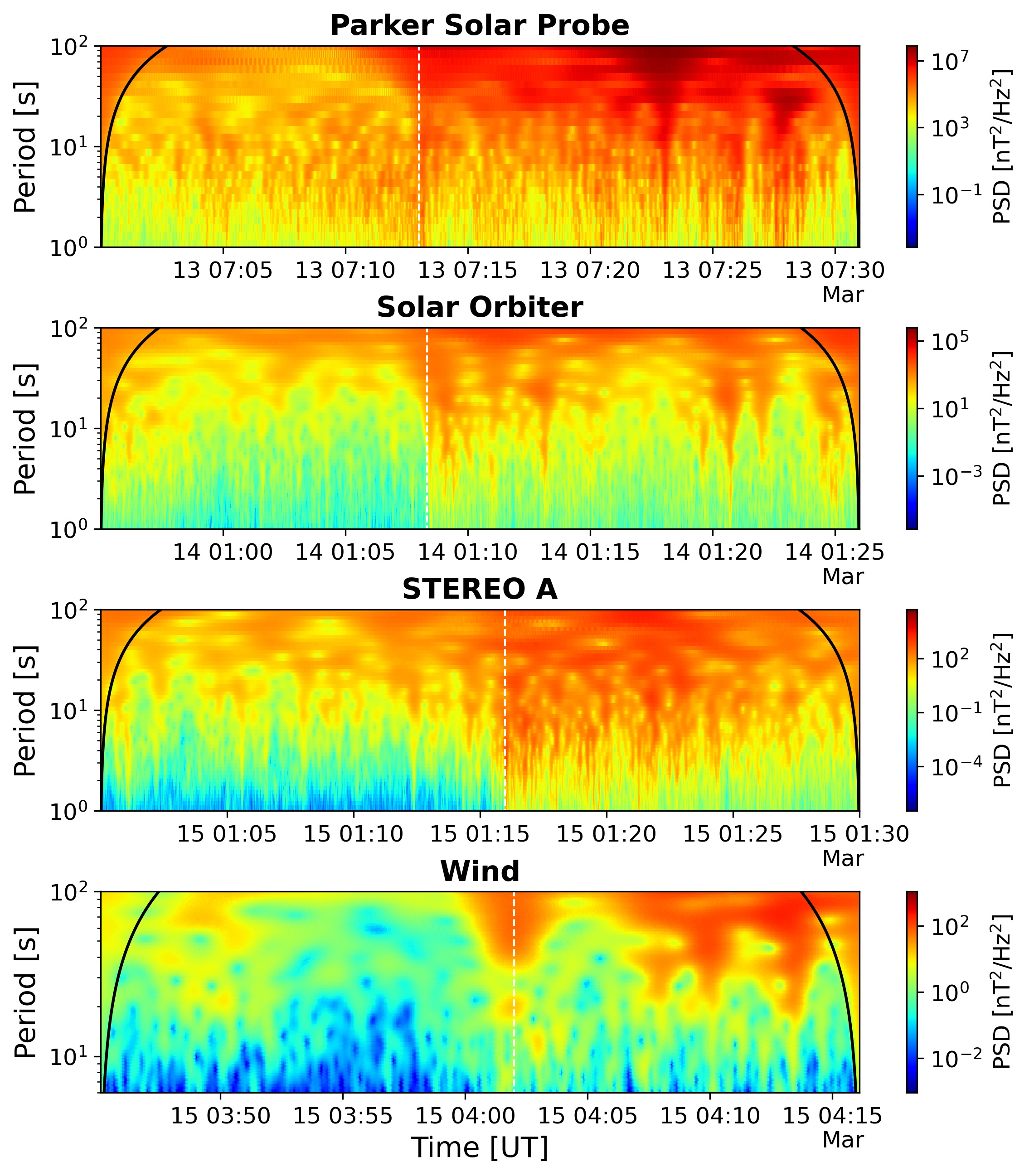}
    \caption{From top to bottom: power spectral density computed using the wavelet coefficients over the three magnetic field components for PSP, Solar Orbiter, STEREO-A, and Wind. In each panel, the vertical dashed line marks the shock crossing time.}
    \label{fig:wavelet}
\end{figure}
In order to characterize the environment through which the shocks propagate, we have studied the properties of the magnetic field turbulence close to the shock front during each satellite crossing. This has been carried out by computing the power spectral density (PSD) of the magnetic field components using the wavelet coefficients \citep{alexandrova_2008}
\begin{equation}
    |\mathcal{W}_{\mathbf{B}}(\tau,t)|^2 = \sum_{i} |\mathcal{W}_i(\tau,t) |^2 ,
\end{equation}
where $\tau = 1/f$ represents a timescale and the sum is over the magnetic field components. $\mathcal{W}_i(\tau,t)$ are the Morlet wavelet coefficients for different timescales $\tau$ and time $t$ \citep{torr-compo_1998},
\begin{equation}
    \mathcal{W}_{i}(\tau,t) = \sum_{j=1}^{N} B_i(t_j)\psi^*\left[ \left( t_j-t \right)/\tau \right] ,
\end{equation}
with $\psi^*$ the conjugate of the wavelet function. 
This enables us to analyze the magnetic energy content both in frequency (or period) and time (as a function of the distance from the shock), and then identify the regions close to the shock front where the magnetic energy is larger.
The PSD as a function of time and time scales is reported in Figure \ref{fig:wavelet} for each shock crossing at PSP, SolO, STEREO-A, and Wind (from top to bottom).  In Figure \ref{fig:wavelet} the scalograms are computed over a time series of about 30 minutes around the shock crossing at each spacecraft, marked by the vertical dashed lines. It is evident how the magnetic field power increases close to the shock and the fluctuations are compressed in the downstream region. Differences in the PSD distribution can be detected at each satellite: in particular a cascade of magnetic energy is observed upstream at PSP and STEREO-A, implying the presence of a highly fluctuating medium over a broad range of time scales (from large to small)-see further discussions in the main text.

\bibliographystyle{aasjournal}

\begin{thebibliography}{}

\bibitem[Ma et al.(2011)]{ma_2011}
Ma, S., Raymond, J.~C., Golub, L., Lin, J., Chen, H., Grigis, P., Testa, P., \& Long, D. (2011).
Observations and interpretation of a low coronal shock wave observed in the EUV by the SDO/AIA.
\textit{The Astrophysical Journal}, \textbf{738}(2), 160.

\bibitem[Trotta et al.(2021)]{Trotta21}
Trotta, D., Valentini, F., Burgess, D., \& Servidio, S. (2021).
Phase space transport in the interaction between shocks and plasma turbulence.
\textit{Proceedings of the National Academy of Science}, \textbf{118}(21), e2026764118.
doi:10.1073/pnas.2026764118.

\bibitem[Mann \& Veronig(2023)]{mann2023}
Mann, G., \& Veronig, A.~M. (2023).
Propagation of a dome-shaped, large-scale extreme-ultraviolet wave in the solar corona.
\textit{Astronomy \& Astrophysics}, \textbf{676}, A144.
doi:10.1051/0004-6361/202245688.

\bibitem[Nisticò et al.(2020)]{Nistico2020}
Nisticò, G., Bothmer, V., Vourlidas, A., Liewer, P.~C., Thernisien, A.~F., Stenborg, G., \& Howard, R.~A. (2020).
Simulating White-Light Images of Coronal Structures for Parker Solar Probe/WISPR: Study of the Total Brightness Profiles.
\textit{Solar Physics}, \textbf{295}(4), 63.
doi:10.1007/s11207-020-01626-y.

\bibitem[Temmer et al.(2013)]{temmer_2013}
Temmer, M., Vršnak, B., \& Veronig, A.~M. (2013).
The Wave–Driver System of the Off-Disk Coronal Wave of 17 January 2010.
\textit{Solar Physics}, \textbf{287}(1), 441–454.

\bibitem[Reames(1999)]{reames_1999}
Reames, D.~V. (1999).
Particle acceleration at the Sun and in the heliosphere.
\textit{Space Science Reviews}, \textbf{90}(3), 413–491.

\bibitem[Desai \& Giacalone(2016)]{desai-giacalone_2016}
Desai, M., \& Giacalone, J. (2016).
Large gradual solar energetic particle events.
\textit{Living Reviews in Solar Physics}, \textbf{13}(1), 3.

\bibitem[Rouillard et al.(2012)]{rouillard_2014}
Rouillard, A.~P., Sheeley, N.~R., Tylka, A., Vourlidas, A., Ng, C.~K., Rakowski, C., Cohen, C.~M.~S., Mewaldt, R.~A., Mason, G.~M., Reames, D., et al. (2012).
The longitudinal properties of a solar energetic particle event investigated using modern solar imaging.
\textit{The Astrophysical Journal}, \textbf{752}(1), 44.

\bibitem[Rouillard et al.(2016)]{rouillard_2016}
Rouillard, A.~P., Plotnikov, I., Pinto, R.~F., Tirole, M., Lavarra, M., Zucca, P., Vainio, R., Tylka, A.~J., Vourlidas, A., De Rosa, M.~L., et al. (2016).
Deriving the properties of coronal pressure fronts in 3D: application to the 2012 May 17 ground level enhancement.
\textit{The Astrophysical Journal}, \textbf{833}(1), 45.

\bibitem[Lario et al.(2016)]{lario_2016}
Lario, D., Kwon, R.-Y., Vourlidas, A., Raouafi, N.~E., Haggerty, D.~K., Ho, G.~C., Anderson, B.~J., Papaioannou, A., Gómez-Herrero, R., Dresing, N., et al. (2016).
Longitudinal properties of a widespread solar energetic particle event on 2014 February 25: evolution of the associated CME shock.
\textit{The Astrophysical Journal}, \textbf{819}(1), 72.

\bibitem[Chiappetta et al.(2021)]{chiappetta_2021}
Chiappetta, F., Laurenza, M., Lepreti, F., \& Consolini, G.\ (2021).
Proton Energy Spectra of Energetic Storm Particle Events and Relation with Shock Parameters and Turbulence.
\textit{The Astrophysical Journal}, \textbf{915}(1), 8. doi:10.3847/1538-4357/abfe09.

\bibitem[Kilpua et al.(2017)]{kilpua_2017}
Kilpua, E., Koskinen, H.~E.~J., \& Pulkkinen, T.~I.\ (2017).
Coronal mass ejections and their sheath regions in interplanetary space.
\textit{Living Reviews in Solar Physics}, \textbf{14}(1), 5. doi:10.1007/s41116-017-0009-6.

\bibitem[Reames(2021)]{reames_2021}
Reames, D.~V.\ (2021).
\textit{Solar Energetic Particles: A Modern Primer on Understanding Sources, Acceleration and Propagation}.
Springer Lecture Notes in Physics, Vol. 978. doi:10.1007/978-3-030-66402-2.

\bibitem[Ontiveros \& Vourlidas(2009)]{ontiveros-vourlidas_2009}
Ontiveros, V., \& Vourlidas, A.\ (2009).
Quantitative measurements of coronal mass ejection-driven shocks from LASCO observations.
\textit{The Astrophysical Journal}, \textbf{693}(1), 267.

\bibitem[Kwon \& Vourlidas(2017)]{kwon-vourlidas_2017}
Kwon, R.-Y., \& Vourlidas, A.\ (2017).
Investigating the wave nature of the outer envelope of halo coronal mass ejections.
\textit{The Astrophysical Journal}, \textbf{836}(2), 246.

\bibitem[Bale et al.(2016)]{Bale_2016SSRv}
Bale, S.~D., Goetz, K., Harvey, P.~R., Turin, P., Bonnell, J.~W., Dudok de Wit, T., Ergun, R.~E., MacDowall, R.~J., Pulupa, M., Andr{\'e}, M., et al.\ (2016).
The FIELDS Instrument Suite for Solar Probe Plus: Measuring the Coronal Plasma and Magnetic Field, Plasma Waves and Turbulence, and Radio Signatures of Solar Transients.
\textit{Space Science Reviews}, \textbf{204}(1–4), 49–82. doi:10.1007/s11214-016-0244-5.

\bibitem[Kasper et al.(2016)]{Kasper_2016SSRv}
Kasper, J.~C., Abiad, R., Austin, G., Balat-Pichelin, M., Bale, S.~D., Belcher, J.~W., Berg, P., Bergner, H., Berthomier, M., Bookbinder, J., et al.\ (2016).
Solar Wind Electrons Alphas and Protons (SWEAP) Investigation: Design of the Solar Wind and Coronal Plasma Instrument Suite for Solar Probe Plus.
\textit{Space Science Reviews}, \textbf{204}(1–4), 131–186. doi:10.1007/s11214-015-0206-3.

\bibitem[Horbury et al.(2020)]{Horbury_2020}
Horbury, T.~S., O'Brien, H., Carrasco Blazquez, I., Bendyk, M., Brown, P., Hudson, R., Evans, V., Oddy, T.~M., Carr, C.~M., Beek, T.~J., et al.\ (2020).
The Solar Orbiter magnetometer.
\textit{Astronomy \& Astrophysics}, \textbf{642}, A9. doi:10.1051/0004-6361/201937257.

\bibitem[Owen et al.(2020)]{Owen_2020}
Owen, C.~J., Bruno, R., Livi, S., Louarn, P., Al Janabi, K., Allegrini, F., Amoros, C., Baruah, R., Barthe, A., Berthomier, M., et al.\ (2020).
The Solar Orbiter Solar Wind Analyser (SWA) suite.
\textit{Astronomy \& Astrophysics}, \textbf{642}, A16. doi:10.1051/0004-6361/201937259.

\bibitem[Lepping et al.(1995)]{Lepping_1995SSRv}
Lepping, R.~P., Acuña, M.~H., Burlaga, L.~F., Farrell, W.~M., Slavin, J.~A., Schatten, K.~H., Mariani, F., Ness, N.~F., Neubauer, F.~M., Whang, Y.~C., et al.\ (1995).
The Wind Magnetic Field Investigation.
\textit{Space Science Reviews}, \textbf{71}(1–4), 207–229. doi:10.1007/BF00751330.

\bibitem[Lin et al.(1995)]{Lin_1995SSRv}
Lin, R.~P., Anderson, K.~A., Ashford, S., Carlson, C., Curtis, D., Ergun, R., Larson, D., McFadden, J., McCarthy, M., Parks, G.~K., et al.\ (1995).
A Three-Dimensional Plasma and Energetic Particle Investigation for the Wind Spacecraft.
\textit{Space Science Reviews}, \textbf{71}(1–4), 125–153. doi:10.1007/BF00751328.

\bibitem[Luhmann et al.(2008)]{Luhmann_2008}
Luhmann, J.~G., Curtis, D.~W., Schroeder, P., McCauley, J., Lin, R.~P., Larson, D.~E., Bale, S.~D., Sauvaud, J.-A., Aoustin, C., Mewaldt, R.~A., et al.\ (2008).
STEREO IMPACT investigation goals, measurements, and data products overview.
In \textit{The STEREO Mission}, Springer, 117–184.

\bibitem[Galvin et al.(2008)]{Galvin_2008SSRv}
Galvin, A.~B., Kistler, L.~M., Popecki, M.~A., Farrugia, C.~J., Simunac, K.~D.~C., Ellis, L., Möbius, E., Lee, M.~A., Boehm, M., Carroll, J., et al.\ (2008).
The Plasma and Suprathermal Ion Composition (PLASTIC) Investigation on the STEREO Observatories.
\textit{Space Science Reviews}, \textbf{136}(1–4), 437–486. doi:10.1007/s11214-007-9296-x.

\bibitem[Paschmann \& Schwartz(2000)]{2000ESASPPaschmann}
Paschmann, G., \& Schwartz, S.~J. (2000).
ISSI Book on Analysis Methods for Multi-Spacecraft Data.
In \textit{Cluster-II Workshop: Multiscale / Multipoint Plasma Measurements}, ESA Special Publication \textbf{449}, 99.

\bibitem[Kwon et al.(2015)]{kwon_2015}
Kwon, R.-Y., Zhang, J., \& Vourlidas, A. (2015).
Are halo-like solar coronal mass ejections merely a matter of geometric projection effects?
\textit{The Astrophysical Journal Letters}, \textbf{799}(2), L29.

\bibitem[van de Hulst(1950)]{vandeHulst_1950}
van de Hulst, H.~C. (1950).
The electron density of the solar corona.
\textit{Bulletin of the Astronomical Institutes of the Netherlands}, \textbf{11}, 135.

\bibitem[Hayes et al.(2001)]{hayes_2001}
Hayes, A.~P., Vourlidas, A., \& Howard, R.~A. (2001).
Deriving the electron density of the solar corona from the inversion of total brightness measurements.
\textit{The Astrophysical Journal}, \textbf{548}(2), 1081.

\bibitem[Leblanc et al.(1998)]{leblanc_1998}
Leblanc, Y., Dulk, G.~A., \& Bougeret, J.-L. (1998).
Tracing the electron density from the corona to 1 AU.
\textit{Solar Physics}, \textbf{183}, 165–180.

\bibitem[Thernisien et al.(2006)]{thernisien_2006}
Thernisien, A.~F.~R., Howard, R.~A., \& Vourlidas, A. (2006).
Modeling of flux rope coronal mass ejections.
\textit{The Astrophysical Journal}, \textbf{652}(1), 763.

\bibitem[Thernisien(2011)]{thernisien_2011}
Thernisien, A. (2011).
Implementation of the graduated cylindrical shell model for the three-dimensional reconstruction of coronal mass ejections.
\textit{The Astrophysical Journal Supplement Series}, \textbf{194}(2), 33.

\bibitem[Bemporad \& Mancuso(2011)]{bemporad_2011}
Bemporad, A., \& Mancuso, S. (2011).
Identification of super- and subcritical regions in shocks driven by coronal mass ejections.
\textit{The Astrophysical Journal Letters}, \textbf{739}(2), L64.

\bibitem[Kwon et al.(2013)]{kwon_2013}
R. Kwon, M. Kramar, T. Wang, L. Ofman, J. M. Davila, J. Chae, J. Zhang,
\textit{Global coronal seismology in the extended solar corona through fast magnetosonic waves observed by STEREO SECCHI COR1},
The Astrophysical Journal, \textbf{776}(1), 55, 2013.

\bibitem[Vr{\v{s}}nak et al.(2002)]{vrvsnak_2002}
B. Vr{\v{s}}nak, J. Magdaleni{\'c}, H. Aurass, G. Mann,
\textit{Band-splitting of coronal and interplanetary type II bursts-II. Coronal magnetic field and Alfv{\'e}n velocity},
Astronomy \& Astrophysics, \textbf{396}(2), 673--682, 2002.

\bibitem[Jebaraj et al.(2024)]{jebaraj_2024}
I. C. Jebaraj, O. V. Agapitov, M. Gedalin, L. Vuorinen, M. Miceli, C. M. S. Cohen, A. Voshchepynets, A. Kouloumvakos, N. Dresing, A. Marmyleva, et al.,
\textit{Direct Measurements of Synchrotron-emitting Electrons at Near-Sun Shocks},
The Astrophysical Journal Letters, \textbf{976}(1), L7, 2024.

\bibitem[Trotta et al.(2024)]{trotta2024}
D. Trotta, H. Hietala, N. Dresing, T. Horbury, Y. Kartavykh, J. Gieseler, I. C. Jebaraj, R. G{\'o}mez-Herrero, F. Espinosa Lara, R. Vainio,
\textit{Solar Orbiter Cycle 25 Interplanetary Shock List}, 2024, Zenodo.

\bibitem[Kwon \& Vourlidas (2018)]{kwon-vourlidas_2018}
R. Kwon, A. Vourlidas,
\textit{The density compression ratio of shock fronts associated with coronal mass ejections},
J. Space Weather Space Clim., \textbf{8}, A08, 2018. DOI: 10.1051/swsc/2017045.

\bibitem[Brueckner et al.(19985)]{brueckner_1995}
G. E. Brueckner, R. A. Howard, M. J. Koomen, C. M. Korendyke, D. J. Michels, J. D. Moses, D. G. Socker, K. P. Dere, P. L. Lamy, A. Llebaria, M. V. Bout, R. Schwenn, G. M. Simnett, D. K. Bedford, C. J. Eyles,
\textit{The Large Angle Spectroscopic Coronagraph (LASCO)},
Solar Physics, \textbf{162}(1-2), 357--402, 1995. DOI: 10.1007/BF00733434.

\bibitem[Kaiser et al.(2008)]{kaiser_2008}
M. L. Kaiser, T. A. Kucera, J. M. Davila, O. C. St. Cyr, M. Guhathakurta, E. Christian,
\textit{The STEREO Mission: An Introduction},
Space Science Reviews, \textbf{136}(1-4), 5--16, 2008. DOI: 10.1007/s11214-007-9277-0.

\bibitem[Vourlidas et al.(2016)]{vourlidas_2016}
A. Vourlidas, R. A. Howard, S. P. Plunkett, C. M. Korendyke, A. F. R. Thernisien, D. Wang, N. Rich, M. T. Carter, D. H. Chua, D. G. Socker, M. G. Linton, J. S. Morrill, S. Lynch, A. Thurn, P. Van Duyne, R. Hagood, G. Clifford, P. J. Grey, M. Velli, P. C. Liewer, J. R. Hall, E. M. DeJong, Z. Mikic, P. Rochus, E. Mazy, V. Bothmer, J. Rodmann,
\textit{The Wide-Field Imager for Solar Probe Plus (WISPR)},
Space Science Reviews, \textbf{204}(1-4), 83--130, 2016. DOI: 10.1007/s11214-014-0114-y.

\bibitem[Kwon et al.(2014)]{kwon_2014}
R. Kwon, J. Zhang, O. Olmedo,
\textit{New insights into the physical nature of coronal mass ejections and associated shock waves within the framework of the three-dimensional structure},
The Astrophysical Journal, \textbf{794}(2), 148, 2014.

\bibitem[Saito et al.(1977)]{saito_1977}
K. Saito, A. I. Poland, R. H. Munro,
\textit{A study of the background corona near solar minimum},
Solar Physics, \textbf{55}, 121--134, 1977.

\bibitem[Dresing et al.(2025)]{dresing_2025}
N. Dresing, I. C. Jebaraj, N. Wijsen, E. Palmerio, L. Rodr{\'\i}guez-Garc{\'\i}a, C. Palmroos, J. Gieseler, M. Jarry, E. Asvestari, J. G. Mitchell, C. M. S. Cohen, C. O. Lee, W. Wei, R. Ramstad, E. Riihonen, P. Oleynik, A. Kouloumvakos, A. Warmuth, B. S{\'a}nchez-Cano, B. Ehresmann, P. Dunn, O. Dudnik, C. Mac Cormack,
\textit{The reason for the widespread energetic storm particle event of 13 March 2023},
Astronomy \& Astrophysics, \textbf{695}, A127, 2025. DOI: 10.1051/0004-6361/202453596.

\bibitem[Alexandrova et al.(2008)]{alexandrova_2008}
O. Alexandrova, V. Carbone, P. Veltri, L. Sorriso-Valvo,
\textit{Small-Scale Energy Cascade of the Solar Wind Turbulence},
The Astrophysical Journal, \textbf{674}(2), 1153--1157, 2008. DOI: 10.1086/524056.

\bibitem[Torrence \& Compo (1998)]{torr-compo_1998}
C. Torrence, G. P. Compo,
\textit{A Practical Guide to Wavelet Analysis},
Bulletin of the American Meteorological Society, \textbf{79}(1), 61--78, 1998. DOI: 10.1175/1520-0477(1998)079<0061:APGTWA>2.0.CO;2.



\end{thebibliography}

\end{document}